\documentclass[12pt]{article}
\usepackage{axodraw}
\usepackage{graphicx}

\newcommand{\sla}[1]{#1\!\!\!\slash}

\newcommand{\MS}{$\overline{\rm MS}$ }
             
\newcommand{\gv}{\mbox{GeV}}

\newcommand{\Lz}{L_z}
\newcommand{\Lw}{L_w}
\newcommand{\Lh}{L_h}
\newcommand{\Lt}{L_t}
\newcommand{\Lb}{L_b}

\newcommand{\lz}{l_z}
\newcommand{\lw}{l_w}
\newcommand{\lh}{l_h}
\newcommand{\lt}{l_t}

\newcommand{\Ltw}{L_{tw}}

\newcommand{\Lwh}{L_{wh}}
\newcommand{\Lwz}{L_{wz}}

\newcommand{\lwh}{l_{wh}}

\newcommand{\htop}{H_t}
\newcommand{\htopMS}{h_t}
\newcommand{\LtMS}{l_t}
\newcommand{\LhMS}{l_h}

\newcommand{\Dh}{\Delta_H}

\catcode`@=11
\newcount\@tempcntc
\def\@citex[#1]#2{\if@filesw\immediate\write\@auxout{\string\citation{#2}}\fi
  \@tempcnta\z@\@tempcntb\m@ne\def\@citea{}\@cite{\@for\@citeb:=#2\do
    {\@ifundefined
       {b@\@citeb}{\@citeo\@tempcntb\m@ne\@citea\def\@citea{,}{\bf
?}\@warning
       {Citation `\@citeb' on page \thepage \space undefined}}%
    {\setbox\z@\hbox{\global\@tempcntc0\csname b@\@citeb\endcsname\relax}%
     \ifnum\@tempcntc=\z@ \@citeo\@tempcntb\m@ne
       \@citea\def\@citea{,}\hbox{\csname b@\@citeb\endcsname}%
     \else
      \advance\@tempcntb\@ne
      \ifnum\@tempcntb=\@tempcntc
      \else\advance\@tempcntb\m@ne\@citeo
      \@tempcnta\@tempcntc\@tempcntb\@tempcntc\fi\fi}}\@citeo}{#1}}
\def\@citeo{\ifnum\@tempcnta>\@tempcntb\else\@citea\def\@citea{,}%
  \ifnum\@tempcnta=\@tempcntb\the\@tempcnta\else
   {\advance\@tempcnta\@ne\ifnum\@tempcnta=\@tempcntb \else
\def\@citea{--}\fi
    \advance\@tempcnta\m@ne\the\@tempcnta\@citea\the\@tempcntb}\fi\fi}
\catcode`@=12

\begin{document}

\title{
\vskip-3cm{\baselineskip14pt
\centerline{\normalsize DESY 14-006\hfill ISSN 0418-9833}
\centerline{\normalsize January 2014\hfill}}
\vskip1.5cm
Two-loop electroweak threshold corrections to the bottom and top Yukawa
couplings}

\author{Bernd A. Kniehl, Oleg L. Veretin
\\
{\normalsize II. Institut f\"ur Theoretische Physik, Universit\"at Hamburg,}
\\
{\normalsize Luruper Chaussee 149, 22761 Hamburg, Germany}
}

\date{}

\maketitle

\begin{abstract}
We study the relationship between the $\overline{\mathrm{MS}}$ Yukawa coupling
and the pole mass for the bottom and top quarks at the two-loop electroweak
order ${\cal O}(\alpha^2)$ in the gaugeless limit of the standard model.
We also consider the $\overline{\mathrm{MS}}$ to pole mass relationships at
this order, which include tadpole contributions to ensure the gauge
independence of the $\overline{\mathrm{MS}}$ masses.
In order to suppress numerically large tadpole contributions, we propose a
redefinition of the running heavy-quark mass in terms of the
$\overline{\mathrm{MS}}$ Yukawa coupling.
We also present $\Delta r$ in the $\overline{\mathrm{MS}}$ scheme at
${\cal O}(\alpha^2)$ in the gaugeless limit.
As an aside, we also list the exact two-loop expression for the mass
counterterms of the bottom and top quarks.

\medskip

\end{abstract}

\newpage

\section{Introduction}

The recent discovery of the Higgs boson \cite{Aad:2012tfa} was a giant leap in
particle physics.
It confirmed that the concept of spontaneous symmetry braking in connection
with the generation of masses by the Higgs mechanism could be realized in
nature.
This does not explain, however, the very large spread of the fermion masses and
the values of the masses themselves.
It is generally believed that some grand unified theories could provide a
solution to this fundamental problem.
An essential r\^ole in such analyses is played by the renormalization group
(RG) equations, which determine the scale dependencies of the running
parameters.  

Due to the large values of their masses, the bottom quark and, even more so,
the top quark attract great interest.
Even disregarding the fact that quark masses require special consideration
because quarks do not appear as free particles, one may introduce different
parameters to describe the notion of quark mass.
The most important definitions are those of the pole mass $M$, the running
mass $m(\mu)$ in the modified minimal-subtraction ($\overline{\rm MS}$)
scheme, and the running Yukawa coupling $y(\mu)$, defined in \MS scheme as
well.
Here, $\mu$ is the 't~Hooft mass of dimensional regularization.
The relationships between these quantities can be obtained in perturbation
theory as series in the strong-coupling constant $\alpha_s$ and Sommerfeld's
fine-structure constant $\alpha$.
The terms containing $\ln\mu^2$ can be obtained using the RG beta functions and
anomalous dimensions.
In QCD, they have been computed in the three- \cite{Tarasov:1980au} and
four-loop \cite{vanRitbergen:1997va} approximations.
In the standard model (SM), the corresponding RG functions, known through two
loops since Ref.~\cite{Fischler:1982du}, have recently been evaluated at the
three-loop level \cite{Mihaila:2012fm}.

The other aspect of the problem is the matching between the running parameters
and the physical observables.
These so-called threshold relationships include not only terms with $\ln\mu^2$,
but also terms of non-logarithmic type.
The relation between the \MS mass and the pole mass of a quark was
elaborated in QCD at one \cite{Tarrach:1980up}, two \cite{Gray:1990yh}, and
three \cite{Chetyrkin:1999ys} loops. 
The two-loop result in the supersymmetric extension of QCD was obtained in
Ref.~\cite{Bednyakov:2002sf}.
These corrections can be readily applied also to the Yukawa coupling of a
quark.
However, the situation becomes more complicated if electroweak corrections are
taken into account.
In this case, the relation between the \MS mass and the \MS Yukawa coupling of
a fermion becomes nontrivial.
The full one-loop corrections to the relationships between the \MS Yukawa
couplings and the pole masses of the fermions were derived in
Ref.~\cite{Hempfling:1994ar}.
The mixed QCD-electroweak two-loop corrections were evaluated for the top quark
in Ref.~\cite{Jegerlehner:2003py} and for the other fermions in
Ref.~\cite{Kniehl:2004hfa} by using the method of large-mass expansion.
In some earlier calculations, the tadpole contributions were omitted in the
\MS to pole mass relations of fermions, e.g.\ in the calculation of the
electroweak parameter $\Delta\rho$ in the gaugeless limit in
Ref.~\cite{Faisst:2003px}. 
Expressed in terms of the top-quark and Higgs-boson pole masses, the result for
$\Delta\rho$ thus obtained is correct, but some intermediate results differ
from those in Ref.~\cite{Jegerlehner:2003py}.
This situation was clarified in Ref.~\cite{Faisst:2004gn}.

The aim of the present paper is to extend the theoretical knowledge of the
relationships between the \MS masses and Yukawa couplings and the pole
masses of the bottom and top quarks by calculating the two-loop electroweak
corrections, at order $O(\alpha^2)$, in the approximation provided by the
gaugeless limit.

This paper is organized as follows.
In Sections~\ref{sec:two} and \ref{sec:three}, we give all the necessary
definitions concerning the \MS mass and Yukawa coupling, respectively.
In Sections~\ref{sec:four} and \ref{sec:five}, we list analytical results for
the bottom and top \MS Yukawa couplings, respectively.
In Section~\ref{sec:six}, we present our numerical analysis.
In Appendix~\ref{app:deltar}, we list two-loop expressions for the electroweak
parameter $\Delta\bar{r}$, which enters the relationships between the \MS
masses and Yukawa couplings, both in the gaugeless and heavy-top-quark limits.
Appendix~\ref{app:renconst} contains the exact two-loop renormalization
constants of the bottom- and top-quark masses in the \MS scheme.
In Appendix~\ref{app:mbMS}, we present the \MS to pole mass relation for the
bottom quark in the heavy-top-quark limit.


\section{Running mass}
\label{sec:two}

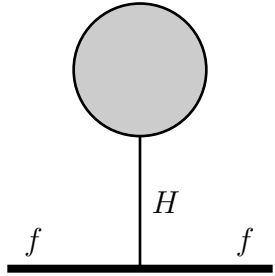
\begin{figure}[ht]
\centerline{
\begin{picture}(100,100)(0,0)
\SetScale{1.0}
\GCirc(50,75){25}{0.8}
\SetWidth{3.0}
\Line(0,0)(100,0)
\SetWidth{1.0}
\Line(50,0)(50,50)
\CArc(50,75)(25,0,360)
\Text(60,25)[c]{$H$}
\Text(10,10)[c]{$f$}
\Text(90,10)[c]{$f$}
\end{picture}
}
\caption{\label{fig:one}%
Tadpole contribution to the self-energy $\Sigma(\sla{p})$ of a fermion $f$.
$H$ stands for the Higgs boson.}
\end{figure}

The pole mass $M$ of a particle is determined by definition from the position
of the pole of its propagator.\footnote{%
If the particle is unstable, then the pole position has a complex value.
In this case, the pole mass is usually taken to be the real part of it.}
We start from the general form of the inverse fermion propagator,
\begin{eqnarray}
\label{invprop}
S^{-1}(\sla{p}) = 
   \sla{p} - m_0 - \Sigma(\sla{p}) \,,
\end{eqnarray}
where $p$ is the four-momentum and $m_0$ is the bare mass of the fermion.
The self-energy function $\Sigma(\sla{p})$ is given by the sum of all
one-particle-irreducible Feynman diagrams pertaining to the transition of the
fermion to itself and depends on further parameters of the theory, such as
masses, couplings, and mixing angles. 
An important comment should be made here.
In the SM with the Higgs mechanism, the inclusion of tadpoles (see
Fig.~\ref{fig:one}) is necessary to render the relationship between the pole
and bare masses gauge independent \cite{Fleischer:1980ub}.
Thus, we include in $\Sigma(\sla{p})$ all Feynman diagrams which are
one-particle irreducible with respect to all particles except for the Higgs
boson.

In the electroweak theory, the left- and right-handed components of a fermion
field propagate differently due to parity violation.
On the other hand, we take the Cabibbo-Kobayashi-Maskawa quark mixing matrix to
be unity and thus effectively turn off CP violation.
In this case, the most general decomposition of $\Sigma(\sla{p})$ reads
\cite{Kniehl:2008cj}:
\begin{eqnarray}
\label{sigma}
\Sigma(\sla{p}) = 
   \sla{p}\,\omega_L A_L(p^2) + \sla{p}\,\omega_R A_R(p^2) + m_0 B(p^2) \,, 
\end{eqnarray}
where $\omega_{L,R} = (1\mp\gamma_5)/2$ are the projectors on the left- and
right-handed spinor components, respectively, and $A_L$, $A_R$, and $B$ are
dimensionless scalar functions of $p^2$ depending also on the SM parameters.

In order to invert the matrix in Eq.~(\ref{invprop}), we first decompose it
into its left- and right-handed components.
The inverse $S(\sla{p})$ may be found in Eq.~(10) of Ref.~\cite{Kniehl:2008cj}.
The poles of its left- and right-handed components coincide and are given by
the solution $p^2=M^2$ of the equation \cite{Kniehl:2008cj}
\begin{eqnarray}
  p^2 [ 1 - A_L(p^2) ] [ 1 - A_R(p^2) ] - m_0^2 [ 1 + B(p^2) ]^2 = 0  \,.
\label{poleeq}
\end{eqnarray}
We can solve Eq.~(\ref{poleeq}) perturbatively by substituting the ansatz
\begin{equation}
M=m_0(1+X_1+X_2+\dots),
\label{eq:m}
\end{equation}
where $X_i$ refers to the order $i$.
Explicitly, through second order, we find
\begin{eqnarray}
\label{Mpole}
  X_1 &=& B_1 + \frac{1}{2} A_{L,1} + \frac{1}{2} A_{R,1} \,,
\nonumber\\
  X_2 &=& B_2 + \frac{1}{2} A_{L,2} + \frac{1}{2} A_{R,2} 
             + X_1(A_{L,1} + A_{R,1} + A_{L,1}' + A_{R,1}' + 2B_1') 
\nonumber\\
&&
           - \frac{1}{2} X_1^2
           - \frac{1}{2} A_{L,1} A_{R,1} 
           + \frac{1}{2} B_1^2 \,.
\end{eqnarray}
In Eq.~(\ref{Mpole}), all functions are to be evaluated at $p^2=m_0^2$ and the
prime stands for the derivative with respect to $p^2/m_0^2$.
The self-energy function in Eq.~(\ref{sigma}) is gauge dependent and, in
general, also infrared divergent.
However, the coefficients $X_i$ in Eq.~(\ref{eq:m}) are manifestly gauge
invariant \cite{Gambino:1999ai}, provided the tadpole diagrams are included,
and infrared safe \cite{Kronfeld:1998di}.
The generalization of Eqs.~(\ref{eq:m}) and (\ref{Mpole}) to the case of
intergeneration mixing is elaborated in Ref.~\cite{Kniehl:2012zb}.

\section{Yukawa coupling}
\label{sec:three}

We start from the SM renormalized in the on-shell scheme, in which Sommerfeld's
fine-structure constant $\alpha=e^2/(4\pi)$ as measured in Thomson scattering
and the pole masses $M_X$ of the elementary particles $X=W,Z,H,t,b,\ldots$
serve as basic parameters, and consider a generic heavy quark with pole mass
$M$.
The sine $S_w$ of the weak-mixing angle and the Yukawa coupling $Y$ of the
heavy quark are derived parameters, defined to all orders as
\begin{equation}
S_w=\sqrt{1-\frac{M_W^2}{M_Z^2}},\qquad Y=2^{-1/2}\frac{eM}{S_wM_W},
\label{eq:derived}
\end{equation}
respectively.
It is advantageous to introduce the Fermi constant $G_F$, which is the
fundamental coupling in the Fermi model of four-fermion interactions, by
equating the total decay width of the muon evaluated in the QED-improved Fermi
model and the SM \cite{Sirlin:1980nh}.
This leads to the relationship
\begin{equation}
2^{5/2}G_F=\frac{e^2}{S_w^2M_W^2}(1+\Delta r),
\label{eq:deltar}
\end{equation}
where $\Delta r$ accommodates all the radiative corrections which the SM
introduces beyond the QED-improved Fermi model \cite{Sirlin:1980nh}.\footnote{%
In the literature, Eq.~(\ref{eq:deltar}) is frequently written with
$(1+\Delta r)$ replaced by $(1-\Delta r)^{-1}$, which is equivalent at one
loop.}
Numerically, we have $G_F=1.16637\times10^{-5}\mbox{GeV}^{-2}$ \cite{pdg}.

We now pass from the original on-shell scheme to the \MS scheme and denote the
counterparts of $e$, $M_W$, $M_Z$, and $M$ as $\bar{e}(\mu)$, $m_W(\mu)$,
$m_Z(\mu)$, and $m(\mu)$, respectively.
Accordingly, Eqs.~(\ref{eq:derived}) and (\ref{eq:deltar}) become
\begin{eqnarray}
s_w(\mu)&=&\sqrt{1-\frac{m_W^2(\mu)}{m_Z^2(\mu)}},\qquad
y(\mu)=2^{-1/2}\frac{\bar{e}(\mu)m(\mu)}{s_w(\mu)m_W(\mu)},
\label{eq:swy}\\
2^{5/2}G_F&=&\frac{\bar{e}^2(\mu)}{s_w^2(\mu)m_W^2(\mu)}[1+\Delta\bar{r}(\mu)],
\label{drdef}
\end{eqnarray}
respectively.
An explicit analytic expression for $\Delta\bar{r}(\mu)$ through two loops may
be found in Appendix~\ref{app:deltar}.\footnote{%
This electroweak parameter and its two-loop corrections significantly differ
from the quantities $\Delta\hat{\rho}$, $\Delta\hat{r}$, and $\Delta\hat{r}_W$
introduced in Ref.~\cite{Sirlin:1989uf} on the basis of a hybrid
renormalization scheme and their two-loop corrections
\cite{Fanchiotti:1992tu}.}
Substituting Eq.~(\ref{drdef}) into the second equality in Eq.~(\ref{eq:swy}),
we find the relationship
\begin{equation}
y(\mu)=2^{3/4}G_F^{1/2}M[1+\delta(\mu)],
\label{eq:yukawa}
\end{equation}
with the radiative correction
\begin{equation}
\delta(\mu)=\frac{m(\mu)}{M}\,\frac{1}{\sqrt{1+\Delta\bar{r}(\mu)}}-1.
\label{eq:threshold}
\end{equation}
In the context of RG analyses, Eq.~(\ref{eq:yukawa}) allows one to relate the
\MS Yukawa coupling $y(\mu)$ at some matching scale $\mu=\mu_0$, which is
typically chosen to be of the order of the Higgs vacuum expectation value
$2^{-1/4}G_F^{-1/2}=246$~GeV, to the physical SM parameters defined in the
on-shell scheme, and Eq.~(\ref{eq:threshold}) is frequently called threshold
correction.

In order for the mass counterterms in the on-shell scheme to be gauge
independent, the tadpole contributions, whose ultraviolet (UV) divergent and
finite parts are both gauge dependent, must be properly included
\cite{Fleischer:1980ub}.
Since the pole and \MS masses are related by the UV-finite parts of these mass
counterterms, the tadpole contributions are indispensable in order for the
gauge independence to be communicated from the pole masses
\cite{Gambino:1999ai} to the \MS masses \cite{Hempfling:1994ar,Kniehl:2004hfa}.
Since the Higgs boson happens to be considerably lighter than the top quark
\cite{Aad:2012tfa}, the UV-finite parts of the tadpole contributions are
numerically sizeable, as was observed for the top quark in
Ref.~\cite{Jegerlehner:2012kn}.
In the case of the bottom quark, they even seriously jeopardize the
perturbative stability of the \MS mass, as will be demonstrated in
Section~\ref{sec:six}.

On the other hand, it was noticed that $\delta(\mu)$ in
Eq.~(\ref{eq:threshold}) is devoid of tadpole contributions in $O(\alpha)$
\cite{Hempfling:1994ar} and $O(\alpha\alpha_s)$ \cite{Kniehl:2004hfa}.
In this paper, we investigate the situation at $O(\alpha^2)$ in the gaugeless
limit.
We find that the tadpoles contained in $m(\mu)/M$ and $\Delta\bar{r}(\mu)$ do
not fully cancel in Eq.~(\ref{eq:threshold}).
While their leading contributions originating from genuine two-loop tadpoles
and products of one-loop tadpoles drop out, subleading contributions arising
from products of one-loop tadpoles and tadpole-free one-loop diagrams survive.
In fact, the latter are indispensable to ensure that the well-known two-loop
expression \cite{Fischler:1982du} for the RG beta function of $y(\mu)$,
\begin{equation}
\beta_y=\frac{\mu^2d}{d\mu^2}y(\mu)
=2^{3/4}G_F^{1/2}M\frac{\mu^2d}{d\mu^2}\delta(\mu),
\label{eq:beta}
\end{equation}
is recovered.
Specifically, the tadpole contributions to $\delta(\mu)$ at $O(\alpha^2)$ in
the gaugeless limit cancel to such a degree that the latter is finite in the
limit of massless Higgs boson.
We verified this also for the full $O(\alpha^2)$ result for $\delta(\mu)$.
In turn, this implies that the sizeable electroweak corrections that challenge
the usefulness of $m(\mu)$ do not plague $\delta(\mu)$.
This provides a strong motivation for us to introduce an alternative definition
of running quark mass by rescaling $y(\mu)$ as\footnote{%
In Ref.~\cite{Jegerlehner:2012kn}, it was denoted as $\hat{m}_f(\mu^2)$.}
\begin{equation}
m_Y(\mu)=2^{-3/4}G_F^{-1/2}y(\mu).
\label{mY}
\end{equation}
We call this the Yukawa mass.
According to Eqs.~(\ref{eq:yukawa}) and (\ref{eq:threshold}), it is related to
the pole and \MS masses as
\begin{equation}
m_Y(\mu)=M[1+\delta(\mu)]
=\frac{m(\mu)}{\sqrt{1+\Delta\bar{r}(\mu)}}.
\label{mym}
\end{equation}
At tree level, we have $m_Y(\mu)=M=m(\mu)$.

If electroweak corrections are neglected and only the strong interactions are
taken into account, then $\Delta\bar{r}(\mu)=0$, so that $m_Y(\mu)=m(\mu)$.
The pure QCD contribution to $\delta(\mu)$ in Eq.~(\ref{eq:threshold}) reads
\cite{Tarrach:1980up,Gray:1990yh}:
\begin{eqnarray}
\label{deltaQCD}
\delta_{\rm QCD}(\mu) &=& 
       \frac{\alpha_s(\mu)}{4\pi} \left( 4 L - \frac{16}{3} \right)
      + \left[ \frac{\alpha_s(\mu)}{4\pi} \right]^2 \left[
                  \left( -14 + \frac{4}{3} n_f \right) L^2
                + \left( \frac{314}{3} - \frac{52}{9} n_f \right) L
\right.
\nonumber\\
&&{}-\left.\frac{3161}{18} 
                - \frac{112}{3} \zeta_2
                - \frac{32}{3} \zeta_2 \ln2
                + \frac{8}{3} \zeta_3
                + \left( \frac{71}{9} + \frac{16}{3} \zeta_2 \right) n_f
          \right] + \mathcal{O}(\alpha_s^3)\,,
\end{eqnarray}
where $\alpha_s(\mu)$ is the strong-coupling constant in the \MS scheme,
$L=\ln(M^2/\mu^2)$, and $n_f$ is the number of quark flavors.
The $\mathcal{O}(\alpha_s^3)$ term in Eq.~(\ref{deltaQCD}) may be found in
Ref.~\cite{Chetyrkin:1999ys}. 

In Sections~\ref{sec:four} and \ref{sec:five}, we shall provide the
electroweak corrections to Eq.~(\ref{eq:threshold}) through two loops for the
bottom and top quarks, respectively.
The corresponding relationships between $m(\mu)$ and $M$ may then be
obtained via Eq.~(\ref{mym}) in connection with the expression for
$\Delta\bar{r}(\mu)$ listed in Appendix~\ref{app:deltar}.

We work in $R_\xi$ gauge with four independent gauge parameters, associated
with the electroweak gauge bosons and the gluon, and perform expansions in
the $\xi$ parameters as explained in
Refs.~\cite{Jegerlehner:2001fb,Jegerlehner:2002em}.
All required \MS renormalization constants may be found through the appropriate
orders in Refs.~\cite{Jegerlehner:2001fb,Jegerlehner:2002em}, except for two,
namely the two-loop renormalization constants for the bottom- and top-quark
masses.
These missing results are presented in exact analytical forms in
Appendix~\ref{app:renconst}.


\section{Bottom quark}
\label{sec:four}

In the case of the bottom quark, the calculation of $\Sigma(\sla{p})$ can
essentially be reduced to the evaluation of vacuum bubble integrals.
By explicit calculation, we find
\begin{eqnarray}
&&\!\!\!\!\!\!\!\!\!\!\!
\frac{m_{{\rm Y},b}(\mu)}{M_b}  =  1
      + \delta_{\rm QCD}(\mu)
\nonumber\\
&&
      + \frac{\alpha}{4\pi S_w^2} \Bigg\{
           \frac{M_t^2}{M_W^2} \Bigg(  - \frac{5}{16} + \frac{3}{8} \Lt 
               + N_c \Big( \frac{1}{8} -\frac{1}{4} \Lt \Big) \Bigg)
          - \frac{3}{8}   \frac{M_W^2}{M_H^2-M_W^2} \Lwh
          + \frac{1}{16} \frac{M_H^2}{M_W^2}
\nonumber\\
&&
          - \frac{3}{8}   \frac{M_W^4}{(M_t^2-M_W^2)^2}   \Ltw
          + \frac{3}{8}   \frac{M_W^2}{M_t^2-M_W^2}  (1 - \Ltw)
          + \frac{3}{8S_w^2} \Lwz
          + S_w^2 Q_b^2 ( -4 + 3 \Lb )
\nonumber\\
&&
          + S_w^2 v_b^2 \Big( -\frac{5}{2}  + 3 \Lz \Big)
          + S_w^2 a_b^2 \Big( -\frac{1}{2}  + 3 \Lz \Big)
          - \frac{5}{8} - \frac{5}{16} \frac{M_Z^2}{M_W^2}
          + \frac{3}{8} \Lw  + \frac{3}{8} \Lh
\nonumber\\
&&
         + \frac{M_b^2}{M_W^2} \Bigg(
              \frac{11}{48}
            + \frac{1}{4} \Lb
            - \frac{1}{8} \Lt
            - \frac{1}{8} \Lz
            - \frac{3}{8} \Lh
\nonumber\\
&&
            + v_b^2 S_w^2 C_w^2 \Big( - \frac{8}{3} - 4 \Lb + 4 \Lz \Big)
              \Bigg)
            +  O\left(\frac{M_b^4}{M_W^4}\right) 
        \Bigg\}
\nonumber\\
&&
      + C_F \frac{\alpha_s}{4\pi} \frac{\alpha}{4\pi S_w^2} \Bigg\{
        \frac{M_t^2}{M_W^2} \Bigg( 
               - \frac{13}{4} - \frac{15}{16} \Lb  - \frac{3}{2} \Lt
               + \frac{9}{8} \Lt \Lb + \frac{9}{8} \Lt^2
\nonumber\\
&&
               + N_c   \Big(
                        \frac{21}{16} - \frac{1}{2} \zeta_2 
                      + \frac{3}{8} (1 - 2 \Lt) \Lb 
                       + \frac{7}{4} \Lt - \frac{3}{4} \Lt^2
                      \Big)
                    \Bigg)
\nonumber\\
&&
       + \Bigg( 3 \frac{M_t^2}{M_W^2}
          + 3
          + \frac{21}{4} \frac{M_W^2}{M_t^2-M_W^2}
          + \frac{9}{4} \frac{M_W^4}{(M_t^2-M_W^2)^2}
            \Bigg) {\rm Li}_2 \Big( 1 - \frac{M_W^2}{M_t^2} \Big)
\nonumber\\
&&
          + \frac{M_W^2}{M_t^2-M_W^2} \Big( -\frac{51}{8} 
               + \frac{9}{8} (1 - \Ltw) \Lb + \frac{39}{8} \Ltw \Big)
          + \frac{M_W^4}{(M_t^2-M_W^2)^2} \Big( \frac{33}{8} 
                    - \frac{9}{8} \Lb \Big)   \Ltw
\nonumber\\
&&
          + \frac{3}{8} \frac{M_W^2}{M_H^2-M_W^2} (4 - 3 \Lb)   \Lwh
          - \frac{1}{16} \frac{M_H^2}{M_W^2} (4 - 3 \Lb)
          - \frac{3}{8S_w^2}   (4 - 3 \Lb)   \Lwz
\nonumber\\
&&
          + S_w^2 Q_b^2 \Big( \frac{7}{4} - 24\zeta_3 - 60\zeta_2 
                + 96 \zeta_2\ln2 - 21\Lb + 9\Lb^2 \Big)
\nonumber\\
&&
          + S_w^2 v_b^2 \Big( \frac{23}{4} - \frac{15}{2}\Lb 
                 - 9\Lz + 9\Lb\Lz \Big)
          + S_w^2 a_b^2 \Big( \frac{55}{4} - \frac{3}{2}\Lb 
                 - 9\Lz - 9\Lb\Lz \Big)
\nonumber\\
&&
          - \frac{23}{16}
          + \frac{5}{4}\frac{M_Z^2}{M_W^2}
          - \frac{15}{16}\frac{M_Z^2}{M_W^2}\Lb
          - \frac{15}{8}\Lb
          + \Big( -\frac{3}{2} + \frac{9}{8}\Lb \Big)
                  \Big( \Lh + \Lw + \frac{M_Z^2}{M_W^2}\Lz \Big)
          - 3\Lw
          + \frac{3}{4}\Lt
            \Bigg\}
\nonumber\\
&&
\nonumber\\
&&
   + \left( \frac{\alpha M_t^2}{4\pi M_W^2 S_w^2} \right)^2
         \left( A_{2,2} \ln^2\frac{M_t^2}{\mu^2}
       + A_{2,1} \ln\frac{M_t^2}{\mu^2} + B_2
    \right)  \,.
\label{mYb}
\end{eqnarray}
Here, $\delta_{\rm QCD}(\mu)$ is given by Eq.~(\ref{deltaQCD}) with $M=M_b$ and
$n_f=5$,
$L_X=\ln(M_X^2/\mu^2)$,
$L_{XY}=L_X-L_Y$,
$C_w=\sqrt{1-S_w^2}$,
$I_b=-1/2$ and $Q_b=-1/3$ are the weak isospin and electric charge of the
bottom quark, and 
$v_b=(I_b - 2Q_b S_w^2)/(2 S_w C_w)$ and
$a_b =I_3/(2 S_w C_w)$ are its vector and axial-vector couplings to the $Z$
boson, respectively.

Equation~(\ref{mYb}) is expanded in powers of $M_b/M_W$, through terms of
$O(M_b^2/M_W^2)$ in $O(\alpha)$ and through terms of $O(M_b^0/M_W^0)$ beyond
that.
In fact, the $O(\alpha M_b^4/M_W^4)$ terms in Eq.~(\ref{mYb}) are already
negligibly small.
The exact analytical result of $O(\alpha)$ may be found in
Ref.~\cite{Hempfling:1994ar}.\footnote{%
Note, however, that the result in Ref.~\cite{Hempfling:1994ar} is written in
terms of $G_F$ and needs to be rewritten in terms of $\alpha$ if it is to be
used in lieu of the $O(\alpha)$ term in Eq.~(\ref{mYb}).}

The $O(\alpha^2)$ term in Eq.~(\ref{mYb}) is calculated adopting the
approximation by the gaugeless limit.
Since its leading behavior in the mass of the heaviest SM particle is of
$O(M_t^4)$, we pull out the factor $M_t^4/M_W^4$ so as to minimize the $M_t$
dependence of the coefficients $A_{2,2}$, $A_{2,1}$, and $B_2$, for which we
obtain
\begin{eqnarray}
\label{ABbottom_gl}
A_{2,2}^{\rm gl} & = &  \frac{63}{128} \,,
\nonumber\\
A_{2,1}^{\rm gl} & = &  - \frac{35}{128} 
          + \left( -\frac{21}{128} + \frac{3}{32} \ln\htop \right) \htop
          - \frac{3}{64}  \htop^2
    + \frac{3}{32} \left( 4 - \htop \right) I_1(\htop)  \,,
\nonumber\\
B_{2}^{\rm gl} &=& \frac{1}{32}(2-\htop)\htop^2 \ln^2\htop
              + \left( -\frac{5}{16} - \frac{1}{16}\htop - \frac{1}{64}\htop^2 \right) \ln\htop
              + \frac{5}{64}(4 - \htop) I_1(\htop)
\nonumber\\
&&
   + \frac{3}{32}(4-\htop) I_2(\htop)
   + \left(
          - \frac{17}{16}
          + \frac{27}{64} \htop
          + \frac{3}{128} \htop^2
          - \frac{1}{64} \htop^3
          \right) \Phi(\htop/4)
\nonumber\\
&&
    +  \left(
          - \frac{93}{64}
          + \frac{11}{16\htop}
          + \frac{3}{4} \htop
          + \frac{7}{64} \htop^2
          - \frac{3}{32} \htop^3
          \right) \Big( {\rm Li}_2(1-\htop) - \zeta_2 \Big) 
          - \frac{5}{32} \htop^3 \zeta_2
\nonumber\\
&&
       + \frac{1451}{512}
          - \frac{13}{8} \zeta_2
       + \Big(
          - \frac{149}{256}
          + \frac{3}{4} \zeta_2
          \Big) \htop
       + \Big(
           \frac{21}{512}
          + \frac{1}{4} \zeta_2
          - \frac{9}{128} S_1
          + \frac{81}{128} S_2
          \Big) \htop^2 \,,\qquad
\end{eqnarray}
where $\htop = M_H^2/M_t^2$,
$S_1=\pi/\sqrt{3}$,
$S_2=4/(9\sqrt{3}){\rm Cl}_2(\pi/3)$
with ${\rm Cl}_2(\theta)$ being Clausen's function defined in Eq.~(\ref{Cl2}),
$\Phi(z)$ is related to a two-loop self-energy integral and defined in
Eq.~(\ref{eq:phi}), and
\begin{eqnarray}
  I_1(z) &=& \int\limits_0^1dt\, \ln[ z(1-t) + t^2]\,,
\nonumber\\
  I_2(z) &=& \int\limits_0^1dt\, \ln\left[\frac{1}{z} - t(1-t) \right] 
\end{eqnarray}
are related to one-loop self-energy integrals and may be expressed in terms of
logarithms.

As is evident from Eq.~(\ref{mY}), the bottom \MS Yukawa coupling
$y_b(\mu)$ simply emerges from Eq.~(\ref{mYb}) by multiplication with the
$\mu$-independent overall factor $2^{3/4}G_F^{1/2}M_b$.
The ratio $m_b(\mu)/M_b$ may be obtained from Eq.~(\ref{mYb}) and the formula
for $\Delta\bar{r}$ given in Appendix~\ref{app:deltar} using Eq.~(\ref{mym}).
For the reader's convenience, we shall present a formula for $m_b(\mu)/M_b$ in
the heavy-top-quark limit in Appendix~\ref{app:mbMS}.
We shall use it in Section~\ref{sec:six} for comparisons with existing
$O(\alpha\alpha_s)$ results and for testing the relevance of the tadpole
contributions at two loops.


\section{Top quark}
\label{sec:five}

The counterpart of Eq.~(\ref{mYb}) for the top quark reads
\begin{eqnarray}
\frac{m_{{\rm Y},t}(\mu)}{M_t} &=&  1 
   + \delta_{\rm QCD}(\mu)
   + \delta_\alpha(\mu)
   + \delta_{\alpha\alpha_s}(\mu)
\nonumber\\
&&
   + \left( \frac{\alpha M_t^2}{4\pi M_W^2 S_w^2} \right)^2
         \left( A_{2,2} \ln^2\frac{M_t^2}{\mu^2}
       + A_{2,1} \ln\frac{M_t^2}{\mu^2} + B_2
    \right)  \,,
\label{mYt}
\end{eqnarray}
where pure QCD correction, $\delta_{\rm QCD}(\mu)$, is given through two loops
by Eq.~(\ref{deltaQCD}) with $M=M_t$ and $n_f=6$,
the one-loop electroweak correction, $\delta_{\alpha}(\mu)$, may be found in
Ref.~\cite{Hempfling:1994ar},
and the mixed two-loop correction $\delta_{\alpha\alpha_s}(\mu)$ was evaluated
in Ref.~\cite{Bezrukov:2012sa}.
Here, we independently obtain this correction by combining $m_t(\mu)/M_t$
evaluated exactly for non-vanishing gauge-boson masses in
Ref.~\cite{Jegerlehner:2003py} with our result for $\Delta\bar{r}$ in
Appendix~\ref{app:deltar} according to Eq.~(\ref{mym}) establishing full
agreement with Ref.~\cite{Bezrukov:2012sa}.
Furthermore, we present here the two-loop electroweak correction, parametrized
by $A_{2,2}$, $A_{2,1}$, and $B_2$.

The case of the top quark is more complicated because a heavy particle now
propagates on external lines.
In such a situation, the result cannot be reduced to vacuum bubble diagrams any
more.
The heavy-top-quark limit leads to non-euclidean expansions, which are rather
involved.
In the gaugeless limit, we find it convenient to introduce the variable
\begin{eqnarray}
  \Dh = 1 - \frac{M_H^2}{M_t^2} \,.
\end{eqnarray}
Assuming the weak neutral scalar resonance recently discovered at the CERN
Large Hadron Collider \cite{Aad:2012tfa} to be the SM Higgs boson, we have
$M_H\approx 125$~GeV, so that $\Dh$ is close to 0.5.
Nevertheless, it appears to be a good expansion parameter.
After the expansion in $\Dh$, the resulting integrals correspond to two-loop
self-energies with one mass and on-shell kinematics.
These integrals may be evaluated with the help of the ONSHELL2 program
\cite{Fleischer:1999tu}.

\begin{figure}[ht]
\begin{picture}(100,100)(0,0)
\SetScale{1.0}
\SetWidth{1.0}
\DashCArc(50,50)(30,0,350){3}
\DashLine(50,20)(50,80){3}
\SetWidth{3.0}
\Line(0,50)(20,50)
\Line(80,50)(100,50)
\CArc(50,50)(30,180,270)
\SetWidth{1.0}
\CArc(50,50)(30,90,180)
\Text(30,15)[c]{$t$}
\Text(30,85)[c]{$H$}
\Text(70,15)[c]{$b$}
\Text(70,85)[c]{$G$}
\Text(58,50)[c]{$G$}
\end{picture}
\hspace*{3mm}
\begin{picture}(100,100)(0,0)
\SetScale{1.0}
\SetWidth{1.0}
\DashCArc(50,50)(30,0,350){3}
\DashLine(50,20)(50,80){3}
\SetWidth{3.0}
\Line(0,50)(20,50)
\Line(80,50)(100,50)
\CArc(50,50)(30,270,360)
\SetWidth{1.0}
\CArc(50,50)(30,0,90)
\Text(30,15)[c]{$b$}
\Text(30,85)[c]{$G$}
\Text(70,15)[c]{$t$}
\Text(70,85)[c]{$H$}
\Text(58,50)[c]{$G$}
\end{picture}
\hspace*{3mm}
\begin{picture}(100,100)(0,0)
\SetScale{1.0}
\SetWidth{1.0}
\DashCArc(50,50)(30,120,60){3}
\DashCArc(50,75)(15,0,360){3}
\SetWidth{3.0}
\Line(0,50)(20,50)
\Line(80,50)(100,50)
\SetWidth{3.0}
\CArc(50,50)(30,0,60)
\CArc(50,50)(30,120,180)
\SetWidth{1.0}
\CArc(50,50)(30,180,360)
\SetWidth{1.0}
\Text(15,70)[c]{$t$}
\Text(85,70)[c]{$t$}
\Text(50,50)[c]{$b$}
\Text(50,30)[c]{$H$}
\Text(50,80)[c]{$G$}
\end{picture}
\hspace*{3mm}
\begin{picture}(100,100)(0,0)
\SetScale{1.0}
\SetWidth{1.0}
\DashCArc(50,50)(30,120,60){3}
\DashCArc(50,75)(15,0,360){3}
\DashCArc(50,50)(30,180,270){3}
\DashCArc(50,50)(30,270,360){3}
\SetWidth{3.0}
\Line(0,50)(20,50)
\Line(80,50)(100,50)
\SetWidth{1.0}
\CArc(50,75)(15,0,180)
\Text(15,70)[c]{$G$}
\Text(85,70)[c]{$G$}
\Text(50,50)[c]{$G$}
\Text(50,30)[c]{$b$}
\Text(50,80)[c]{$H$}
\end{picture}
\caption{\label{fig:two}%
Four Feynman diagrams whose expansions in $\Dh$ suffer from threshold
singularities.
$G_0$ and $G$ denote the neutral and the charged Goldstone bosons,
respectively.
Their masses vanish in the gaugeless limit.
}
\end{figure}
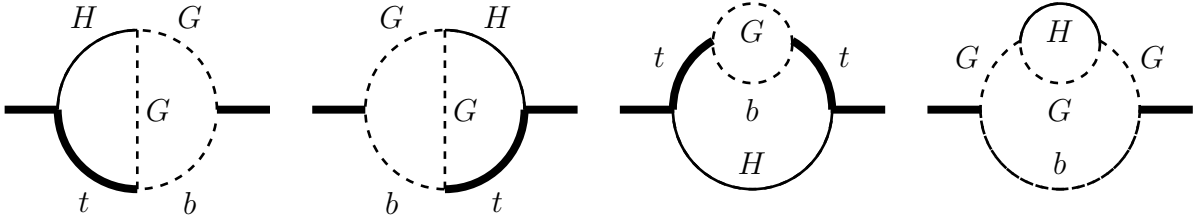

There is, however, one problem related to this procedure.
In fact, such a na\"\i ve expansion can break down if certain threshold
singularities are present.
In the case under consideration, this occurs, for example, when a Feynman
diagram has a unitary cut intersecting one Higgs boson line and one or two
massless lines.
Taken off-shell, such diagrams produce terms of the type
$(p^2-M_H)^n \ln(p^2-M_H^2)$, which disappear in the limit $p^2\to M_H^2$.
The presence of such terms tells us that the expansion in the variable
$p^2-M_H^2$ breaks down at some order $n$.
This issue may be discussed in connection with pole masses of gauge bosons on
the basis of the results obtained in Ref.~\cite{Jegerlehner:2002em}.
Detailed analysis reveals that the first five coefficients of the na\"\i ve
expansion yield the correct result.
In the present case, we find that the first six coefficients of the expansion in
$\Dh$ give the correct result, but, starting from $O(\Dh^6)$, threshold
singularities appear.
In dimensional regularization, they manifest themselves as poles in
$\varepsilon=2-d/2$ with $d$ being the space-time dimensionality, which are not
compensated by renormalization.
Detailed inspection reveals that, among the more than two hundred diagrams
contributing to $\Sigma(\sla{p})$, only four are plagued by threshold
singularities.
They are shown in Fig.~\ref{fig:two}.
These diagrams were evaluated without expansions, using the results for the
master integrals from Ref.~\cite{Jegerlehner:2002em}. 

For the coefficients $A_{2,2}$, $A_{2,1}$, and $B_2$ in Eq.~(\ref{mYt}), we
obtain
\begin{eqnarray}
A_{2,2}^{\rm gl} & = & \frac{243}{128} \,,
\\
A_{2,1}^{\rm gl} & = & -\frac{333}{128}+\frac{81}{64}S_1
     - \frac{117}{128} \Dh 
     + \Big( \frac{3}{8} - \frac{27}{64} S_1 \Big) \Dh^2 
     + \Big( \frac{9}{64} - \frac{3}{16} S_1 \Big) \Dh^3
     + \Big( \frac{9}{256} - \frac{3}{32} S_1 \Big)
\nonumber\\
&&\times \Dh^4
     + \Big( \frac{9}{640} - \frac{1}{16} S_1 \Big) \Dh^5 
     + \Big( \frac{3}{640} - \frac{13}{288} S_1 \Big) \Dh^6 
     + \Big( \frac{3}{4480} - \frac{5}{144} S_1 \Big) \Dh^7 
\nonumber\\
&&
     + \Big( -\frac{3}{2240} - \frac{1}{36} S_1 \Big) \Dh^8 
     + \Big( -\frac{3}{1280} - \frac{89}{3888} S_1 \Big) \Dh^9
     +  O(\Dh^{10}) \,,
\label{A21}\\
 B_2^{\rm gl} &=&
       - \frac{2099}{512}
          - \zeta_3
          + \frac{557}{192} \zeta_2
          - \frac{19}{32} S_1
          + \frac{405}{128} S_2
          + \frac{81}{16} S_2 S_1
       + \Dh   \Big(
           \frac{259}{256}
          + \frac{1}{4} \zeta_3
          - \frac{103}{48} \zeta_2
\nonumber\\
&&
          + \frac{567}{128} S_1
          - \frac{1107}{64} S_2
          - \frac{9}{4} S_2 S_1
          \Big)
       + \Dh^2   \Big(
           \frac{637}{512}
          + \frac{1}{4} \zeta_3
          - \frac{2033}{384} \zeta_2
          + \frac{1151}{192} S_1
          - \frac{3051}{256} S_2
\nonumber\\
&&
          - \frac{9}{16} S_2 S_1
          \Big)
       + \Dh^3   \Big(
           \frac{47}{288}
          - \frac{2765}{576} \zeta_2
          + \frac{18925}{3456} S_1
          - \frac{987}{128} S_2
          \Big)
       + \Dh^4   \Big(
           \frac{2755}{4608}
          - \frac{5275}{1152} \zeta_2
\nonumber\\
&&
          + \frac{913}{192} S_1
          - \frac{1537}{256} S_2
          \Big)
       + \Dh^5   \Big(
           \frac{3697}{4608}
          - \frac{2713}{576} \zeta_2
          + \frac{239501}{51840} S_1
          - \frac{3283}{640} S_2
          \Big)
\nonumber\\
&&
       + \Dh^6   \Big(
           \frac{368489}{414720}
          - \frac{19789}{4320} \zeta_2
          + \frac{1018007}{233280} S_1
          - \frac{26267}{5760} S_2
          - \frac{1}{768} \ln\Dh
          \Big)
\nonumber\\
&&       + \Dh^7   \Big(
           \frac{9679259}{9525600}
          - \frac{47041}{10080} \zeta_2
          + \frac{14055353}{3265920} S_1
          - \frac{125141}{30240} S_2
          - \frac{9}{4480} \ln\Dh
          \Big)
\nonumber\\
&&
       + \Dh^8   \Big(
           \frac{1314795901}{1219276800}
          - \frac{47467}{10368} \zeta_2
          + \frac{36031201}{8709120} S_1
          - \frac{26375}{6912} S_2
          - \frac{127}{53760} \ln\Dh
          \Big)
\nonumber\\
&&
       + \Dh^9   \Big(
           \frac{6467713271}{5486745600}
          - \frac{144331}{31104} \zeta_2
          + \frac{289007057}{70543872} S_1
          - \frac{1549721}{435456} S_2
          - \frac{607}{241920} \ln\Dh
          \Big)
\nonumber\\
&&
       + O(\Dh^{10}) \,.
\label{B2}
\end{eqnarray}
We observe that, in Eq.~(\ref{B2}), the coefficients of the expansion in $\Dh$
contain terms proportional to $\ln\Dh$ starting from $O(\Dh^6)$.
This is how the threshold singularities mentioned above manifest themselves.
Fortunately, this does not spoil the convergence property of Eq.~(\ref{B2}).
In fact, for $M_H=125$~GeV, the terms of $O(\Dh^n)$ with $n=0,\dots,9$ in
Eq.~(\ref{B2}) add up as
\begin{eqnarray}
B_{2}^{\rm gl} &\approx& 1.6088 + 0.1120 + 0.0786 + 0.0213 +  0.0068 + 0.0025 
+ 0.0010 + 0.0004
\nonumber\\
&&{}  + 0.0002 + 0.0001
\nonumber\\
&=& 1.8397 \,,
\end{eqnarray}
exhibiting rapid convergence.
For comparison, we also display the convergence property of the power series in
Eq.~(\ref{A21}), which is not challenged by $\ln\Dh$ terms:
\begin{eqnarray}
A_{2,1}^{\rm gl} &\approx& -0.3060 - 0.4396 - 0.0903 - 0.02219 - 0.0072 
- 0.0026 - 0.0010
\nonumber\\
&&{}- 0.0004 -  0.0001
\nonumber\\ 
&=& -0.8691  \,.
\end{eqnarray}


\section{Discussion}
\label{sec:six}

We are now in a position to present our numerical analysis.
For this, we adopt the following values for the input parameters from
Ref.~\cite{pdg}:
\begin{eqnarray}
M_Z &=& 91.1876~\gv\,, 
\quad 
M_W = 80.385~\gv\,,
\quad 
M_b = 4.89~\gv\,,
\quad 
M_t = 173.5~\gv\,,\footnotemark
\nonumber \\ 
G_F &=&1.16637\times 10^{-5}~\gv^{-2}\,, 
\quad
\alpha^{-1} = 137.035999\,,
\quad 
\alpha_s^{(5)}(M_Z) = 0.1184\,.
\label{params}
\end{eqnarray}
\footnotetext{%
The values of the top-quark mass quoted by the experimental collaborations
correspond to parameters in Monte Carlo event generators in which, apart from
parton showering, the partonic subprocesses are calculated at the tree level,
so that a rigorous theoretical definition of the top-quark mass is lacking
\cite{pdg,Abazov:2011pta}.
For definiteness, we take the value from Ref.~\cite{pdg} to be the pole mass
$M_t$.}
Furthermore, we take the effective fine-structure constant at the scale of the
$Z$-boson mass to be $\alpha^{-1}(M_Z) = 127.944$.
We neglect the masses $M_f$ of all light fermions $f\ne b,t$, since their
effects are negligible and do not play any r\^ole in our considerations.

Through the three-loop order, the QCD relation between the \MS and pole
masses is given by \cite{Chetyrkin:1999ys}
\begin{eqnarray}
[m_t(M_t) - M_t]_{\rm QCD}&=& 
M_t 
\Biggl[ 
- \frac{4}{3}\, \frac{\alpha_s^{(6)}(M_t)}{\pi}\, 
- 9.125\left( \frac{\alpha_s^{(6)}(M_t)}{\pi} \right)^2
\nonumber\\
&&{}- 80.405 \left( \frac{\alpha_s^{(6)}(M_t)}{\pi} \right)^3    
\Biggr] \,.
\end{eqnarray}
These corrections carry over to the mass difference $m_{Y,t}(M_t)-M_t$ because
$\Delta\bar{r}$ in Eq.~(\ref{mym}) does not receive pure QCD corrections, as is
evident from Eq.~(\ref{drexp}).
Using $\alpha_s^{(6)}(M_t)=0.1079$ \cite{Kniehl:2012rz}, which follows from
the value of $\alpha_s^{(5)}(M_Z)$ in Eq.~(\ref{params}) via four-loop
evolution and three-loop matching \cite{Chetyrkin:1997sg}, we obtain the
numerical result 
\begin{eqnarray}
[m_t(M_t) - M_t]_{\rm QCD}= 
 - 7.95~\gv 
 - 1.87~\gv
 - 0.57~\gv
 = -10.38~\gv \,.
\label{qcd:num}
\end{eqnarray}

Let us now estimate how well the approximations by the gaugeless limit work
for the relationships of the \MS masses $m(\mu)$ and Yukawa couplings $y(\mu)$
to the pole masses $M$ for the bottom and top quarks.
Since, at two loops, the exact SM results are not yet available for comparison,
we are led to do this at one loop, using the exact results from
Ref.~\cite{Hempfling:1994ar}.
Let us first consider the \MS masses.
In the gaugeless limit, the one-loop corrections are given by
\begin{eqnarray}
\frac{m_b(\mu)}{M_b} & = & 1
   + \frac{\alpha M_t^2}{4\pi M_W^2 S_w^2} \Bigg\{
    \left( - \frac{N_c}{\htop} + \frac{3}{8} + \frac{3 \htop}{8} \right) \ln\frac{M_t^2}{\mu^2}
       + \frac{N_c}{\htop} - \frac{5}{16} - \frac{3 \htop}{8} 
\nonumber\\
&&{}
     + \frac{3 \htop}{8} \ln\htop  \Bigg\} + \cdots\,,
\nonumber\\ 
\frac{m_t(\mu)}{M_t} & = & 1
   + \frac{\alpha M_t^2}{4\pi M_W^2 S_w^2} \Bigg\{
    \left( - \frac{N_c}{\htop} - \frac{3}{8} + \frac{3 \htop}{8} \right) \ln\frac{M_t^2}{\mu^2}
       + \frac{N_c}{\htop} + 1 - \frac{\htop}{2} 
\nonumber\\
 &&{}  
       +  \frac{\htop^2}{16} \ln\htop
       - \frac{\htop^2}{8} \left(\frac{4}{\htop} - 1 \right)^{3/2}
         \arccos\left( \frac{\sqrt{\htop}}{2} \right)
    \Bigg\} + \cdots \,,
\label{oneloopgl}
\end{eqnarray}
where $\htop$ is defined below Eq.~(\ref{ABbottom_gl}).
With the input parameters in Eq.~(\ref{params}), we obtain the following
values (in GeV) in the gaugeless limit (g.l.) and the full SM (full):
\begin{eqnarray}
[m_b(\mu) - M_b]_{O(\alpha)} &=& \left \{
           { -2.93 + 0.62\times \ln(\mu[\mbox{GeV}]) \qquad \mbox{g.l.}
             \atop 
             -3.06 + 0.66\times \ln(\mu[\mbox{GeV}]) \qquad \mbox{full} } \right.\,,
\label{eq:glb}\\
{[m_t(\mu) - M_t]}_{O(\alpha)} &=& \left \{
           {-123.85 + 26.52\times \ln(\mu[\mbox{GeV}]) \qquad \mbox{g.l.}
             \atop 
            - 118.92 + 25.35\times \ln(\mu[\mbox{GeV}]) \qquad \mbox{full} }\right.\,.
\label{eq:glt}
\end{eqnarray}
Evaluating Eq.~(\ref{eq:glb}) at $\mu=M_b$, we find shifts of $-1.95$~GeV and
$-2.02$~GeV, differing by only 3\%.
Similarly, Eq.~(\ref{eq:glt}) at $\mu=M_t$ yields shifts of 12.91~GeV and
11.78~GeV, with 9\% difference. 
We thus conclude that the gaugeless limit provides a reasonable approximation
for the \MS masses of the bottom and top quarks.
This may be partially traced to the fact that the major contributions arise
from the top-quark tadpole, which is preserved by the gaugeless limit. 

\begin{table}[h!]
\centering
\caption{The various loop contributions to $m_t(M_t)-M_t$ in GeV.}
\label{tab:results}
$\begin{array}{|c|c|c|c|c|c|}
\hline
M_H~[\mbox{GeV}] & {\rm QCD} & O(\alpha)  & O(\alpha \alpha_s) &
       O(\alpha^2) &
\mbox{Total}  \\ \hline
124 & $-10.38$ & 12.11 & $-0.39$ & $-0.48$ & 0.83 \\ \hline
125 & $-10.38$ & 11.91 & $-0.39$ & $-0.46$ & 0.65 \\ \hline
126 & $-10.38$ & 11.71 & $-0.38$ & $-0.44$ & 0.46 \\ \hline
\end{array}$
\end{table}

A detailed analysis of the relationship $m_t(M_t)-M_t$, including the
contributions of orders $O(\alpha_s^n)$ with $n=1,2,3$ and
$O(\alpha\alpha_s^n)$ with $n=0,1$, has recently been presented in
Ref.~\cite{Jegerlehner:2012kn}.
We are now in the position to extend this analysis to order $O(\alpha^2)$.
Table~\ref{tab:results} corresponds to Table~1 in Ref.~\cite{Jegerlehner:2012kn}
with the corresponding column added.
For $M_H=125$~GeV, the $O(\alpha^2)$ shift in $m_t(M_t)-M_t$
estimated in the gaugeless limit is about $-0.5$~GeV, bringing $m_t(M_t)$
even closer to $M_t$.
In view of the above discussion, we expect that this shift will receive a
correction of order 10\% once the residual $O(\alpha^2)$ terms will become
available.
The $O(\alpha_s^4)$ QCD correction to $m_t(M_t)-M_t$ was estimated in
Ref.~\cite{Kataev:2010zh} to be about 20~MeV.
It is plausible to assign a theoretical uncertainty of order 100~MeV, i.e.\
twice the magnitude of the $O(\alpha^2)$ shift, to the values in the
rightmost column in Table~\ref{tab:results}.

The situation is different for the bottom quark.
For $\mu=M_b$, the shifts of orders $O(\alpha)$, $O(\alpha\alpha_s)$, and
$O(\alpha^2)$ in $m_b(M_b)-M_b$ read
\begin{equation}
[m_b(M_b) - M_b]_{O(\alpha,\alpha\alpha_s,\alpha^2)} 
= -1.94~\mbox{GeV} - 1.56~\mbox{GeV} + 1.61~\mbox{GeV} \,,
\label{eq:msmb}
\end{equation}
respectively.
We observe that the two-loop electroweak correction as estimated in the
gaugeless limit is uncomfortably large.
Based on the one-loop consideration above, we expect the error due to the
lack of knowledge of the residual $O(\alpha^2)$ terms to be of order 5\%. 
The abnormal size of the $O(\alpha^2)$ term in Eq.~(\ref{eq:msmb}) is related
to the tadpole contribution as will become apparent below.

\begin{figure}[h!]
\label{plot}
\begin{center}
\centerline{\includegraphics[width=0.8\textwidth,bb=1 1 360 222,clip]{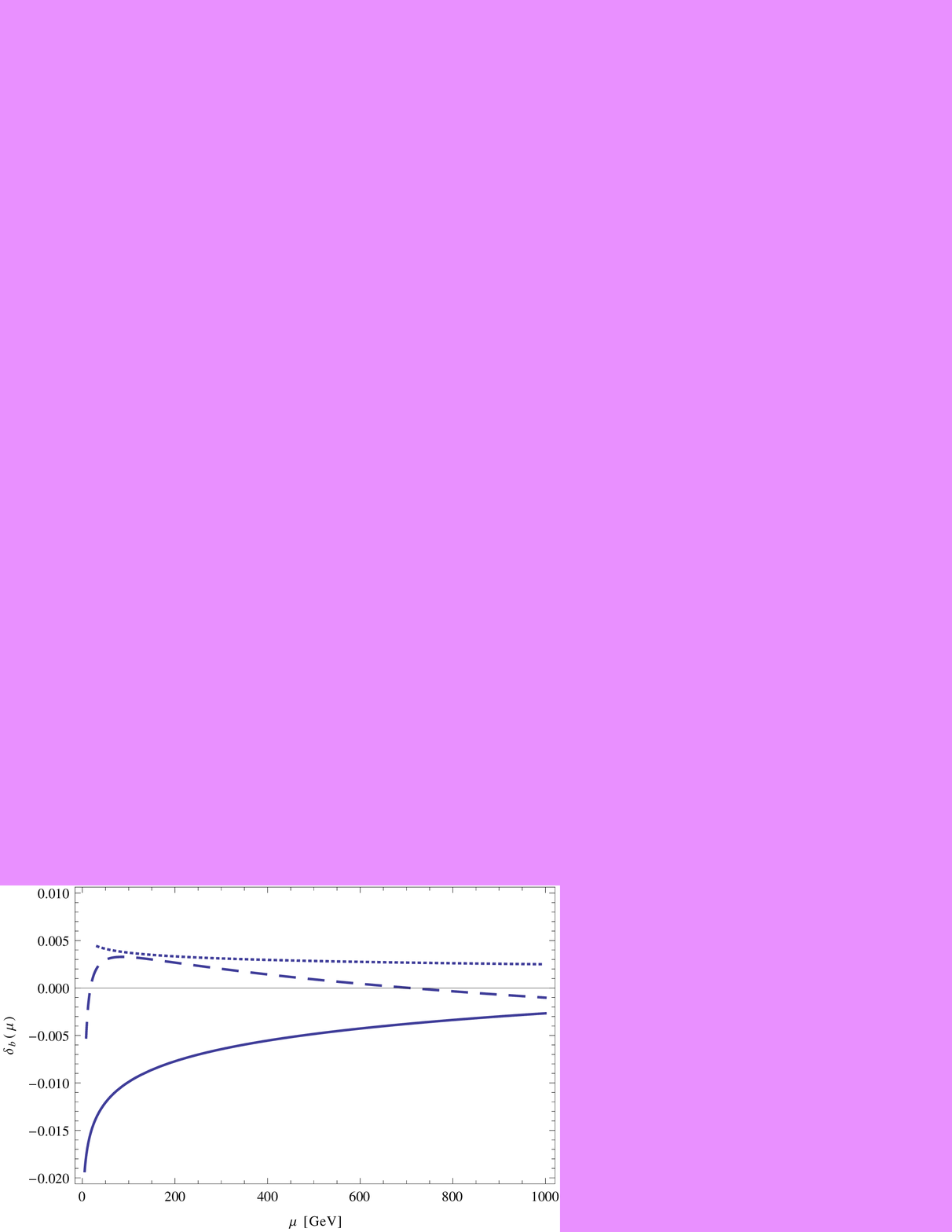}}
\vspace{-6mm}
\centerline{(a)\hfill}
\vspace{6mm}
\centerline{\includegraphics[width=0.8\textwidth,bb=1 1 360 227,clip]{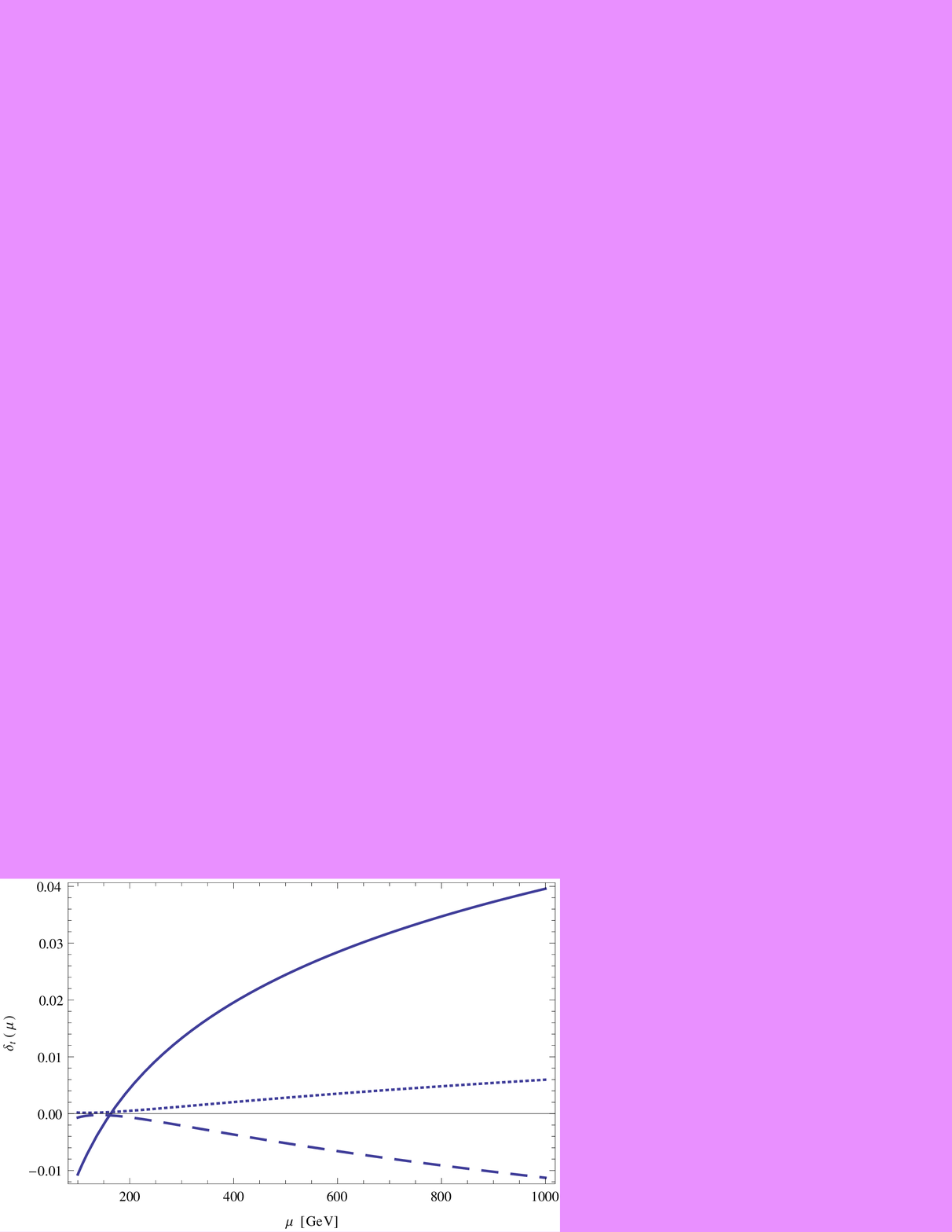}}
\vspace{-6mm}
\centerline{(b)\hfill}
\vspace{6mm}
\end{center}
\caption{\label{fig:three}
Corrections of orders $O(\alpha)$ (solid lines), $O(\alpha\alpha_s)$
(long-dashed lines), and $O(\alpha^2)$ (short-dashed lines) to (a)
$m_{Y,b}(\mu)/M_b$ and (b) and $m_{Y,t}(\mu)/M_t$ given in Eqs.~(\ref{mYb}) and
(\ref{mYt}), respectively, as functions of $\mu$.
The $O(\alpha^2)$ corrections are evaluated in the gaugeless-limit
approximation.
The QCD corrections are not shown.}
\end{figure}
We now turn to the \MS Yukawa couplings or, equivalently, the Yukawa masses
$m_Y(\mu)$.
In Figs.~\ref{fig:three}(a) and (b), we display the corrections of orders
$O(\alpha)$, $O(\alpha\alpha_s)$, and $O(\alpha^2)$ to $m_{Y,b}(\mu)/M_b$ and
$m_{Y,t}(\mu)/M_t$ given in Eqs.~(\ref{mYb}) and (\ref{mYt}), respectively, as
functions of $\mu$.
We observe that, in the case of the bottom quark, the $O(\alpha^2)$ correction
exceeds the $O(\alpha\alpha_s)$ one in size over a large $\mu$ range, while,
in the case of the top quark, the $O(\alpha^2)$ correction is always much
smaller than the $O(\alpha\alpha_s)$ one.
At the thresholds $\mu=M_b,M_t$, where Eqs.~(\ref{mYb}) and (\ref{mYt}) act as
matching conditions for the RG evolution of the \MS Yukawa couplings, we
have\footnote{%
In Eq.~(2.49) of Ref.~\cite{Buttazzo:2013uya}, a numerical interpolation
formula for $y_t(\mu)$ including also the $O(\alpha^2)$ correction may be
found.
Unfortunately, the $O(\alpha)$ and $O(\alpha^2)$ corrections are not
distinguished there.
Since their sum is greatly dominated by the $O(\alpha)$ correction, the
$O(\alpha^2)$ one cannot be extracted reliably enough to allow for any
meaningful comparison with our result in Eq.~(\ref{mYt}).}
\begin{eqnarray}
\frac{m_{Y,b}(M_b)}{M_b} 
         &=& 1+\delta_{\rm QCD}(M_b) - 0.0197 - 0.0113 + 0.0057 \,,
\label{deltab}
\\
\frac{m_{Y,t}(M_t)}{M_t}
         &=& 1+\delta_{\rm QCD}(M_t) + 0.0013 - 0.0004 +  0.0003 \,,
\label{deltat}
\end{eqnarray}
where the last three numbers on the right-hand sides represent the $O(\alpha)$,
$O(\alpha\alpha_s)$, and $O(\alpha^2)$ corrections.
In contrast to $m(\mu)$, $m_Y(\mu)$ is devoid of leading tadpole contributions
as per construction in Eq.~(\ref{mym}).
As a consequence, the electroweak corrections to $m_Y(\mu)/M$ are considerably
smaller in size than those to $m(\mu)/M$, as is evident from the
comparisons of Eq.~(\ref{eq:msmb}) with Eq.~(\ref{deltab}) for the bottom quark
and of Table~\ref{tab:results} with Eq.~(\ref{deltat}) for the top quark.
By the same token, the gaugeless-limit approximation is expected to be less
reliable for $m_Y(\mu)/M$ than for $m(\mu)/M$, in the sense that its
relative deviation from the full SM result is larger, while the absolute
deviation may be similar.
In the remainder of this section, we investigate this issue quantitatively,
using again the one-loop results as benchmarks. 
From the full-SM and gaugeless-limit formulas in Ref.~\cite{Hempfling:1994ar},
we obtain
\begin{eqnarray}
\label{yb}
\left[\frac{m_{Y,b}(\mu)}{M_b}-1\right]_{O(\alpha)} &=& \left \{
           { -0.0490 + 0.0097\times \ln(\mu[\mbox{GeV}]) \qquad \mbox{g.l.}
             \atop 
             -0.0244 + 0.0031\times \ln(\mu[\mbox{GeV}]) \qquad \mbox{full} } \right.
\,,\\
\label{yt}
\left[\frac{m_{Y,t}(\mu)}{M_t}-1\right]_{O(\alpha)} &=& \left \{
           {- 0.141 + 0.029\times \ln(\mu[\mbox{GeV}]) \qquad \mbox{g.l.}
             \atop 
            - 0.111 + 0.022\times \ln(\mu[\mbox{GeV}]) \qquad \mbox{full} }  \right.
\,.
\end{eqnarray}
At first sight, the agreement still seems to be good, especially for the top
quark.
However, evaluating Eq.~(\ref{yb}) at $\mu=M_b$, we obtain $-0.0336$ versus
$-0.0195$, i.e.\ a relative deviation of more than 70\%, and 
evaluating Eq.~(\ref{yt}) at $\mu=M_t$, we obtain 0.0098 versus 0.0013, i.e.\ a
relative deviation of more than 700\%.
Obviously, the $O(\alpha^2)$ corrections displayed in Fig.~\ref{fig:three} have
to be taken with a grain of salt.
While they allow us to estimate the residual theoretical uncertainties more
reliably, their numerical values are only indicative.
We conclude that the electroweak perturbative expansions of $m_Y(\mu)/M$ for
the bottom and top quarks exhibit useful
convergence behaviors, while the $O(\alpha^2)$ terms require more work.


\section{Conclusion}
\label{sec:seven}

In this paper, we calculated the electroweak corrections to the relationships
between the \MS masses $m(\mu)$ and Yukawa couplings
$y(\mu)$ of the bottom and top quarks and their pole masses $M$ at two
loops in the gaugeless limit of the SM.

We verified at one loop that the gaugeless limit provides a reliable
approximation for the \MS masses $m(\mu)$.
In the top-quark case, the new $O(\alpha^2)$ correction induces a shift of
about $-0.5$~GeV in mass difference $m_t(M_t)-M_t$ and reduces its
theoretical uncertainty down to the order of 100~MeV.
In the bottom-quark case, however, the new $O(\alpha^2)$ correction is nearly
as large in size as the $O(\alpha)$ one, indicating that the convergence
property of the perturbative expansion is jeopardized by the tadpole
contributions.

Detailed inspection revealed that the bulk of the corrections to the mass
difference $m(M)-M$ originates from the tadpole contributions, which
are shared by the quantity $\Delta\bar{r}(\mu)$ defined in
Eq.~(\ref{drdef}).
Owing to cancellations on the right-hand side of Eq.~(\ref{mym}), the
relationship between $y(\mu)$ and $M$ is devoid of leading tadpole
contributions, which motivated our proposition to measure the running of the
heavy-quark masses by using $y(\mu)$ after appropriate rescaling, as in
Eq.~(\ref{mY}).
We called this new mass parameter Yukawa mass $m_Y(\mu)$.
In fact, the electroweak corrections to the mass difference $m_Y(M)-M$ are
typically one order of magnitude smaller than those to $m(M)-M$.
While this considerably improves the convergence properties of the
electroweak perturbation expansions, especially for the bottom quark, it
greatly reduces the usefulness of the gaugeless-limit approximation, as we
demonstrated at one loop.
In the top-quark case, we found the $O(\alpha^2)$ correction thus estimated
to be small against the $O(\alpha)$ one, but comparable to the
$O(\alpha\alpha_s)$ one.
In the bottom-quark case, the hierarchy among the $O(\alpha)$,
$O(\alpha\alpha_s)$, and $O(\alpha^2)$ corrections turned out as na\"\i vely
expected.

\section*{Acknowledgements}

The authors thank M.Yu.~Kalmykov for fruitful discussions.
This work was supported in part by the German Federal Ministry of Education
and Research BMBF through Grant No.\ 05H12GUE and by the German Research
Foundation DFG through the Collaborative Research Centre No.\ SFB~676
{\it Particles, Strings and the Early Universe: the Structure of Matter and
Space-Time}.

%
%

\appendix

\boldmath
\section{$\Delta\bar{r}(\mu)$ in the $\overline{\rm MS}$ scheme}
\label{app:deltar}
\unboldmath

In this appendix, we present the electroweak parameter $\Delta\bar{r}$
appearing in Eq.~(\ref{drdef}) through two loops in the SM.
In the following, all masses and couplings are the running quantities defined
in the \MS scheme.
To simplify the notation, the argument $\mu$ is always omitted.
We thus need to specify the coefficients $x_{1,0}$, $x_{1,1}$, and $x_{2,0}$ in
the perturbative expansion
\begin{equation}
\Delta\bar{r} = 
       \frac{g^2}{16\pi^2} \, x_{1,0} 
     + C_F \frac{g_s^2}{16\pi^2} \frac{g^2}{16\pi^2} \, x_{1,1}  
     + \left( \frac{g^2}{16\pi^2} \right)^2 x_{2,0} + \cdots  \,,
\label{drexp}
\end{equation}
where $g=\bar{e}/s_w$ and $g_s$ are the electroweak and strong-coupling
constants in the \MS scheme.

The coefficients $x_{1,0}$ and $x_{1,1}$ may be given in exact form:
\begin{eqnarray}
x_{1,0}  &=&
          N_c \frac{2 m_t^4}{m_W^2 m_H^2} ( 1 - \lt )
        - N_c \frac{m_t^2}{m_W^2} \Big( \frac{1}{4} - \frac{1}{2} \lt \Big)
        + \frac{3}{4} \frac{m_W^2}{m_H^2-m_W^2} \ln\frac{m_W^2}{m_H^2}
\nonumber\\
&&
        + \frac{m_H^2}{m_W^2} \Big( - \frac{7}{8} + \frac{3}{4} \lh \Big)
        + \frac{m_W^2}{m_H^2} \Big( - 1 - \frac{1}{2c_w^4} 
              + 3 \lw + \frac{3}{2c_w^4} \lz \Big)
\nonumber\\
&&
          + \frac{5}{4}
          + \frac{5}{8c_w^2}
          - \frac{3}{4s_w^2} \ln c_w^2
          - \frac{3}{4} \lh
          - \frac{3}{4} \lw
          - \frac{3}{4c_w^2} \lz  \,,
\nonumber\\
x_{1,1}  &=&
         N_c\frac{m_t^4}{m_W^2 m_H^2} ( 20 - 20 \lt + 12 \lt^2 )
         + N_c \frac{m_t^2}{m_W^2} \Big( -\frac{13}{8} + \zeta_2 + \lt 
              - \frac{3}{2} \lt^2 \Big) \,,
\end{eqnarray}
where, in analogy with the definitions introduced below Eq.~(\ref{mYb}),
$l_x=\ln(m_x^2/\mu^2)$, $c_w=m_W/m_Z$, and $s_w=\sqrt{1-c_w^2}$.

The two-loop electroweak correction $x_{2,0}$ in the full SM with arbitrary
particle masses may be expressed in terms of dilogarithms. 
However the result is too lengthy to be presented here.
Instead, we evaluate $x_{2,0}$ in two different approximations:
the gaugeless limit of the SM, $x_{2,0}^{\rm gl}$, and the heavy-top-quark
limit, $x_{2,0}^{\rm ht}$. 
In the former case, we have
\begin{eqnarray}
x_{2,0}^{\rm gl} &=&
             \frac{4 N_c^2 m_t^8}{m_H^4 m_W^4} (1 - \LtMS)
           + \frac{N_c m_t^6}{m_H^2 m_W^4}  \Bigg\{
                           \frac{N_c}{2} (1 - \LtMS) (-3 + 2 \LtMS)
                           - \frac{45}{32} \LtMS^2 
                           + \frac{3}{32} \LhMS^2 
\nonumber\\
&&
                           - \frac{3}{16} \LtMS \LhMS 
                           + \frac{13}{2} \LtMS 
                           - \frac{5}{8} \zeta_2
                           - \frac{15}{2} 
                           - \frac{3}{16} {\cal H}(\htopMS)
                           + \frac{9}{2} \Phi(\htopMS/4)
                         \Bigg\} 
\nonumber\\
&&
           + N_c \frac{m_t^4}{m_W^4}\Bigg\{
                        \Big( \frac{141}{64} - \frac{3}{16} \htopMS - \frac{3}{64} \htopMS^2 
                                 + \frac{N_c}{4} \Big)   \LtMS^2
                      + \Big( -\frac{15}{64} + \frac{3}{16} \htopMS - \frac{3}{64} \htopMS^2 \Big) \LhMS^2
\nonumber\\
&&
                      + \Big( -\frac{81}{32} + \frac{3}{4} \htopMS + \frac{3}{32} \htopMS^2 \Big)  \LtMS\LhMS
                      + \Big( -\frac{5}{4} - \frac{11}{16} \htopMS - \frac{N_c}{4} \Big) \LtMS
\nonumber\\
&&
                      + \Big( \frac{43}{8} - \frac{25}{16} \htopMS \Big)   \LhMS
                      + \frac{N_c}{16}
                      + \frac{7}{16} \zeta_2
                       - \frac{295}{64} + \frac{17}{8} \htopMS
\nonumber\\
&&
                      + \Big( \frac{15}{32} - \frac{3}{8} \htopMS + \frac{3}{32} \htopMS^2 \Big)
                                    {\cal  H}(\htopMS)
                      + \Big( -\frac{15}{8} - \frac{3}{16} \htopMS + \frac{3}{32} \htopMS^2 \Big)
                           \Phi(\htopMS/4)
                      \Bigg\}
\nonumber\\
&&
           + \frac{m_H^4}{m_W^4}  \Bigg\{  
                        \frac{27}{32} \LhMS^2
                      - \frac{27}{8} \LhMS
                      + \frac{1}{4} \zeta_2
                      - \frac{243}{32} S_2
                      + \frac{457}{128}
                     \Bigg\}  \,,
\label{eq:x20}
\end{eqnarray}
where $\htopMS = m_H^2/m_t^2$.
In the latter case, we obtain
\begin{eqnarray}
x_{2,0}^{\rm ht}  
      &=&  N_c \Bigg\{
          N_c \frac{4 m_t^8}{m_W^4 m_H^2} (1 - \lt)
          + \frac{m_t^6}{m_W^4 m_H^2} \Bigg( - \frac{15}{2} - \zeta_2 
                   + \frac{13}{2} \lt - \frac{3}{2} \lt^2
\nonumber\\
&&
                    + \frac{N_c}{2} (1 - \lt) ( -3 + 2 \lt )  \Bigg)
          + \frac{m_t^4}{(m_H^2-m_W^2)^2} \Big( \frac{3}{4} \lh^2 - \frac{3}{4} \lw^2 
                   + 3 \lwh \lt \Big)
\nonumber\\
&&
          + \frac{m_t^4}{m_W^2(m_H^2-m_W^2)} \Big(
                    \frac{3}{2} \lw (1-\lt) 
                     + \frac{3}{2} (\lt - \lh) + \frac{3}{8}(\lt + \lh)^2
                                \Bigg)
\nonumber\\
&&
          + \frac{m_t^4}{m_H^4} \Big(
                   - 8 - \frac{4}{c_w^4} + 6 \lw + 6 \lt + \frac{3}{c_w^4} ( \lt + \lz )
                          \Big)
\nonumber\\
&&
          + \frac{m_t^4}{m_W^2 m_H^2} \Bigg(
                   \frac{4}{3} \frac{s_w^2}{c_w^2} \lt^2
                    + (1-\lt) \Big(
                          \frac{59}{18} + \frac{179}{36c_w^2} - \frac{3}{2s_W^2} \ln c_w^2
                         - \frac{3}{2c_w^2} \lz - 3 \lw \Big)
                            \Bigg)
\nonumber\\
&&
          + \frac{m_t^4}{m_W^4} \Bigg(
                     N_c \Big( \frac{1}{16} - \frac{1}{4} \lt (1-\lt) \Big)
                    + \frac{305}{64} + \frac{11}{8} \zeta_2 + \frac{1}{2} \lh 
                    + \frac{9}{8} \lh^2
\nonumber\\
&&
                    - \frac{21}{4} \lh \lt + \frac{29}{8} \lt + \frac{57}{16} \lt^2
                           \Bigg)
               \Bigg\} + O(m_t^2) \,.
\end{eqnarray}
In Eq.~(\ref{eq:x20}), two new functions have been introduced.
They correspond to two-loop vacuum bubble integrals with two different masses,
namely $J(0,m_1^2,m_2^2)$ and $J(m_1^2,m_2^2,m_2^2)$.
Adopting the notations from Ref.~\cite{Berends:1994sa} with $z=m_1^2/m_2^2$ and
from Ref.~\cite{Davydychev:1992mt} with $z=m_1^2/(4m_2^2)$, they are given by
\begin{eqnarray}
  {\cal H}(z) &=& 2{\rm Li}_2(1 - z) + \frac{1}{2}\ln^2 z \,,
\\
  \Phi(z) &=& 4 \sqrt{\frac{z}{1-z}} {\rm Cl}_2(2\arcsin\sqrt{z}),
\label{eq:phi}
\end{eqnarray}
respectively, where
\begin{equation}
   {\rm Cl}_2(\theta) = - \int\limits_0^\theta  d\theta^\prime
             \ln\left(2\sin\frac{\theta^\prime}{2}\right) 
\label{Cl2}
\end{equation}
is Clausen's integral \cite{polylogs}.


\section{Renormalization constants of the quark masses}
\label{app:renconst}

The calculations presented in this paper require the renormalizations of the
bottom- and top-quark masses through two loops and of all other masses and
couplings at one loop.
For our purposes, knowledge of the corresponding renormalization constants in
the approximations of the gaugeless and heavy-top-quark limits is sufficient.
However, we also calculated the mass renormalization constants $Z_m$ of the
bottom and top quarks exactly at two loops in the SM.
For future applications and checks, we present our results in the following.
In the \MS scheme, $Z_m$ is defined as
\begin{equation}
Z_m=\left[\frac{m^0}{m(\mu)}\right]^2,
\end{equation}
and may be written in the form
\begin{eqnarray}
Z_m &=& 1 
        + \frac{g^2}{16\pi^2} \frac{Z_{\alpha}^{(1,1)}}{\varepsilon}
        + C_F \frac{g_s^2}{16\pi^2} \frac{Z_{\alpha_s}^{(1,1)}}{\varepsilon}
\nonumber\\
&&
        + C_F \frac{g^2}{16\pi^2} \frac{g_s^2}{16\pi^2} 
           \left( \frac{Z_{\alpha\alpha_s}^{(2,2)}}{\varepsilon^2}
                  +\frac{Z_{\alpha\alpha_s}^{(2,1)}}{\varepsilon} \right)
        + \left( \frac{g^2}{16\pi^2} \right)^2
           \left( \frac{Z_{\alpha^2}^{(2,2)}}{\varepsilon^2}
                  +\frac{Z_{\alpha^2}^{(2,1)}}{\varepsilon} \right)
        + \cdots  \,.
\end{eqnarray}
Using the same notation as in Appendix~\ref{app:deltar}, we have
\begin{eqnarray}
Z^{(1,1)}_\alpha &=&
      -\frac{1}{3} 
     - \frac{3}{4} \frac{m_H^2}{m_W^2} 
     - \frac{3}{4} \frac{m_t^2}{m_W^2} 
     - 3 \frac{m_W^2}{m_H^2 }
     + \frac{1}{3} \frac{m_Z^2}{m_W^2} 
     -  \frac{3}{2} \frac{m_Z^4}{m_H^2 m_W^2} 
     + 2 N_c \frac{m_t^4}{m_H^2 m_W^2} \,,
\\
Z^{(1,1)}_{\alpha_s} &=& - 3 \,,
\\
Z^{(2,2)}_{\alpha\alpha_s} &=& 
       2 
     + \frac{9}{2} \frac{m_H^2}{m_W^2} 
     + \frac{27}{4} \frac{m_t^2}{m_W^2} 
     + 18 \frac{m_W^2}{m_H^2 }
     - 2 \frac{m_Z^2}{m_W^2 }
     + 9 \frac{m_Z^4}{m_H^2 m_W^2} 
     -  24 N_c \frac{m_t^4}{m_H^2 m_W^2} \,,
\\
Z^{(2,1)}_{\alpha\alpha_s} &=& 
       \frac{25}{12} 
     - 3 \frac{m_t^2}{m_W^2 }
     + \frac{31}{24} \frac{m_Z^2}{m_W^2}
     + 4 N_c \frac{m_t^4}{m_H^2 m_W^2} \,,
\\
Z^{(2,2)}_{\alpha^2} &=& 
        \frac{83}{24}
          + \frac{9}{16} \frac{m_H^2 m_t^2}{m_W^4}
          - \frac{9}{32} \frac{m_H^4}{m_W^4}
          + \frac{9}{32} \frac{m_Z^2 m_t^2}{m_W^4}
          + \frac{5}{16} \frac{m_Z^2 m_H^2}{m_W^4}
          + \frac{9}{8} \frac{m_Z^4}{m_W^4} \frac{m_t^2}{m_H^2}
\nonumber\\
&&{}
          + \frac{85}{48} \frac{m_Z^4}{m_W^4}
          - \frac{15}{8} \frac{m_Z^6}{m_W^4 m_H^2}
          + \frac{9}{4} \frac{m_Z^8}{m_W^4 m_H^4}
          + \frac{9}{16} \frac{m_t^2}{m_W^2}
          + \frac{11}{8} \frac{m_H^2}{m_W^2}
          - \frac{1}{6} \frac{m_Z^2}{m_W^2}
\nonumber\\
&&{}
          - \frac{5}{4} \frac{m_Z^4}{m_W^2 m_H^2}
          + \frac{9}{4} \frac{m_t^2}{m_H^2}
          + \frac{29}{4} \frac{m_Z^2}{m_H^2}
          + 9 \frac{m_Z^4}{m_H^4}
          + 18 \frac{m_W^2}{m_H^2}
          + 9 \frac{m_W^4}{m_H^4}
\nonumber\\
&&{}
       + n_G   \Bigg(
           \frac{1}{6}
          + \frac{1}{6} \frac{m_Z^4}{m_W^4}
          - \frac{3}{2} \frac{m_Z^6}{m_W^4 m_H^2}
          - \frac{1}{3} \frac{m_Z^2}{m_W^2}
          + 3 \frac{m_Z^4}{m_W^2 m_H^2}
          - 2 \frac{m_Z^2}{m_H^2}
          - \frac{m_W^2}{m_H^2}
          \Bigg)
\nonumber\\
&&{}
       + N_c   \Bigg(
          - \frac{39}{16} \frac{m_t^4}{m_W^4}
          - \frac{3}{8} \frac{m_H^2 m_t^2}{m_W^4}
          - \frac{2}{3} \frac{m_Z^2 m_t^4}{m_W^4 m_H^2}
          - 6 \frac{m_Z^4 m_t^4}{m_W^4 m_H^4}
          + \frac{3}{4} \frac{m_Z^4 m_t^2}{m_W^4 m_H^2}
\nonumber\\
&&{}
          + \frac{2}{3} \frac{m_t^4}{m_W^2 m_H^2}
          - 12 \frac{m_t^4}{m_H^4}
          + \frac{3}{2} \frac{m_t^2}{m_H^2}
          \Bigg)
       + N_c n_G   \Bigg(
           \frac{11}{162}
          + \frac{11}{162} \frac{m_Z^4}{m_W^4}
          - \frac{11}{18} \frac{m_Z^6}{m_W^4 m_H^2}
\nonumber\\
&&{}
          - \frac{11}{81} \frac{m_Z^2}{m_W^2}
          + \frac{11}{9} \frac{m_Z^4}{m_W^2 m_H^2}
          - \frac{10}{9} \frac{m_Z^2}{m_H^2}
          - \frac{m_W^2}{m_H^2}
          \Bigg)
       + 4 N_c^2 \frac{m_t^8}{m_W^4 m_H^4}  \,,
\\
Z^{(2,1)}_{\alpha^2} &=& 
        \frac{103}{144}
          + \frac{11}{32} \frac{m_t^4}{m_W^4}
          + \frac{33}{32} \frac{m_H^4}{m_W^4}
          - \frac{79}{192} \frac{m_Z^2 m_t^2}{m_W^4}
          - \frac{3}{8} \frac{m_Z^2 m_H^2}{m_W^4}
          - \frac{1019}{576} \frac{m_Z^4}{m_W^4}
          + \frac{59}{24} \frac{m_Z^6}{m_W^4 m_t^2}
\nonumber\\
&&{}
          + \frac{53}{96} \frac{m_t^2}{m_W^2}
          - \frac{3}{4} \frac{m_H^2}{m_W^2}
          + \frac{4}{9} \frac{m_Z^2}{m_W^2}
          + \frac{31}{12} \frac{m_Z^4}{m_W^2 m_H^2}
          - \frac{17}{2} \frac{m_Z^2}{m_H^2}
          - \frac{176}{3} \frac{m_W^2}{m_H^2}
\nonumber\\
&&{}
       + n_G   \Bigg(
          - \frac{37}{72}
          - \frac{47}{144} \frac{m_Z^4}{m_W^4}
          + 2 \frac{m_Z^6}{m_W^4 m_H^2}
          + \frac{47}{72} \frac{m_Z^2}{m_W^2}
          - 4 \frac{m_Z^4}{m_W^2 m_H^2}
          + \frac{8}{3} \frac{m_Z^2}{m_H^2}
          + \frac{4}{3} \frac{m_W^2}{m_H^2}
          \Bigg)
\nonumber\\
&&{}
       + N_c   \Bigg(
          - \frac{5}{2} \frac{m_t^6}{m_W^4 m_H^2}
          - \frac{11}{32} \frac{m_t^4}{m_W^4}
          + \frac{3}{8} \frac{m_H^2 m_t^2}{m_W^4}
          + \frac{4}{9} \frac{m_Z^2 m_t^4}{m_W^4 m_H^2}
          + \frac{19}{12} \frac{m_Z^4}{m_W^4} \frac{m_t^2}{m_H^2}
\nonumber\\
&&{}
          - \frac{4}{9} \frac{m_t^4}{m_W^2 m_H^2}
          - \frac{20}{3} \frac{m_Z^2 m_t^2}{m_W^2 m_H^2}
          + \frac{35}{6} \frac{m_t^2}{m_H^2}
          \Bigg)
       + N_c n_G   \Bigg(
          - \frac{623}{1944}
          - \frac{517}{3888} \frac{m_Z^4}{m_W^4}
\nonumber\\
&&{}
          + \frac{22}{27} \frac{m_Z^6}{m_W^4 m_H^2}
          + \frac{517}{1944} \frac{m_Z^2}{m_W^2}
          - \frac{44}{27} \frac{m_Z^4}{m_W^2 m_H^2}
          + \frac{40}{27} \frac{m_Z^2}{m_H^2}
          + \frac{4}{3} \frac{m_W^2}{m_H^2}
          \Bigg) \,,
\end{eqnarray}
for the bottom quark and
\begin{eqnarray}
Z^{(1,1)}_\alpha &=&
       \frac{2}{3} 
     - \frac{3}{4} \frac{m_H^2}{m_W^2} 
     + \frac{3}{4} \frac{m_t^2}{m_W^2} 
     - 3 \frac{m_W^2}{m_H^2 }
     - \frac{2}{3} \frac{m_Z^2}{m_W^2} 
     -  \frac{3}{2} \frac{m_Z^4}{m_H^2 m_W^2} 
     + 2 N_c \frac{m_t^4}{m_H^2 m_W^2}  \,,
\\
Z^{(1,1)}_{\alpha_s} &=& - 3 \,,
\\
Z^{(2,2)}_{\alpha\alpha_s} &=&
         - 4  
         + 18 \frac{m_W^2}{m_H^2} 
         + 4 \frac{m_Z^2}{m_W^2} 
         + 9 \frac{m_Z^4}{m_H^2 m_W^2}  
         + \frac{9}{2} \frac{m_H^2}{m_W^2} 
         - \frac{27}{4} \frac{m_t^2}{m_W^2}
         - 24 N_c \frac{m_t^4}{m_H^2 m_W^2} \,,
\label{eq:z22}\\
Z^{(2,1)}_{\alpha\alpha_s} &=&
       \frac{31}{12} 
     + 3 \frac{m_t^2}{m_W^2 }
     + \frac{19}{24} \frac{m_Z^2}{m_W^2}
     + 4 N_c \frac{m_t^4}{m_H^2 m_W^2} \,,
\label{eq:z21}\\
Z^{(2,2)}_{\alpha^2} &=& 
            \frac{85}{24}
          + \frac{9}{16} \frac{m_t^4}{m_W^4}
          - \frac{9}{16} \frac{m_H^2 m_t^2}{m_W^4}
          - \frac{9}{32} \frac{m_H^4}{m_W^4}
          - \frac{33}{32} \frac{m_t^2 m_Z^2}{m_W^4}
          + \frac{17}{16} \frac{m_Z^2 m_H^2}{m_W^4}
          - \frac{9}{8} \frac{m_Z^2 m_t^4}{m_W^4 m_H^2}
\nonumber\\
&&{}
          + \frac{89}{48} \frac{m_Z^4}{m_W^4}
          - \frac{3}{8} \frac{m_Z^6}{m_W^4 m_H^2}
          + \frac{9}{4} \frac{m_Z^8}{m_W^4 m_H^4}
          + \frac{3}{16} \frac{m_t^2}{m_W^2}
          + \frac{5}{8} \frac{m_H^2}{m_W^2}
\nonumber\\
&&{}
          - \frac{1}{3} \frac{m_Z^2}{m_W^2}
          - \frac{11}{4} \frac{m_Z^4}{m_W^2 m_H^2}
          - \frac{9}{4} \frac{m_t^2}{m_H^2}
          + \frac{41}{4} \frac{m_Z^2}{m_H^2}
          + 9 \frac{m_Z^4}{m_H^4}
          + 15 \frac{m_W^2}{m_H^2}
          + 9 \frac{m_W^4}{m_H^4}
\nonumber\\
&&{}
        + n_G   \Bigg(
          - \frac{1}{3}
          - \frac{1}{3} \frac{m_Z^4}{m_W^4}
          - \frac{3}{2} \frac{m_Z^6}{m_W^4 m_H^2}
          + \frac{2}{3} \frac{m_Z^2}{m_W^2}
          + 3 \frac{m_Z^4}{m_W^2 m_H^2}
          - 2 \frac{m_Z^2}{m_H^2}
          - \frac{m_W^2}{m_H^2}
          \Bigg)
\nonumber\\
&&{}
       + N_c   \Bigg(
            3 \frac{m_t^6}{m_W^4 m_H^2}
          - \frac{33}{16} \frac{m_t^4}{m_W^4}
          - \frac{3}{8} \frac{m_H^2 m_t^2}{m_W^2}
          - \frac{8}{3} \frac{m_Z^2 m_t^4}{m_W^4 m_H^2}
          - 6 \frac{m_Z^4 m_t^4}{m_W^4 m_H^4}
          + \frac{3}{4} \frac{m_Z^4 m_t^2}{m_W^4 m_H^2}
\nonumber\\
&&{}
          + \frac{8}{3} \frac{m_t^4}{m_W^2 m_H^2}
          - 12 \frac{m_t^4}{m_H^4}
          + \frac{3}{2} \frac{m_t^2}{m_H^2}
          \Bigg)
       + N_c n_G   \Bigg(
          - \frac{11}{81}
          - \frac{11}{81} \frac{m_Z^4}{m_W^4}
          - \frac{11}{18} \frac{m_Z^6}{m_W^4 m_H^2}
\nonumber\\
&&{}
          + \frac{22}{81} \frac{m_Z^2}{m_W^2}
          + \frac{11}{9} \frac{m_Z^4}{m_W^2 m_H^2}
          - \frac{10}{9} \frac{m_Z^2}{m_H^2}
          - \frac{m_W^2}{m_H^2}
          \Bigg)
       + 4 N_c^2 \frac{m_t^8}{m_W^4 m_H^4} \,,
\\
Z^{(2,1)}_{\alpha^2} &=& 
          + \frac{11}{48}
          + \frac{3}{16} \frac{m_t^4}{m_W^4}
          - \frac{3}{8} \frac{m_t^2 m_H^2}{m_W^4}
          + \frac{33}{32} \frac{m_H^4}{m_W^4}
          + \frac{223}{192} \frac{m_Z^2 m_t^2}{m_W^4}
          - \frac{3}{8} \frac{m_Z^2 m_H^2}{m_W^4}
          - \frac{289}{192} \frac{m_Z^4}{m_W^4}
\nonumber\\
&&{}
          + \frac{59}{24} \frac{m_Z^6}{m_W^4 m_H^2}
          + \frac{91}{96} \frac{m_t^2}{m_W^2}
          - \frac{3}{4} \frac{m_H^2}{m_W^2}
          + \frac{2}{3} \frac{m_Z^2}{m_W^2}
          + \frac{31}{12} \frac{m_Z^4}{m_W^2 m_H^2}
          - \frac{17}{2} \frac{m_Z^2}{m_H^2}
          - \frac{176}{3} \frac{m_W^2}{m_H^2}
\nonumber\\
&&{}
         + n_G   \Bigg(
          - \frac{7}{72}
          + \frac{13}{144} \frac{m_Z^4}{m_W^4}
          + 2 \frac{m_Z^6}{m_W^4 m_H^2}
          - \frac{13}{72} \frac{m_Z^2}{m_W^2}
          - 4 \frac{m_Z^4}{m_W^2 m_H^2}
          + \frac{8}{3} \frac{m_Z^2}{m_H^2}
          + \frac{4}{3} \frac{m_W^2}{m_H^2}
          \Bigg)
 \nonumber\\
&&{}
          + N_c  \Bigg(
          - \frac{5}{2} \frac{m_t^6}{m_W^4 m_H^2}
          - \frac{25}{32} \frac{m_t^4}{m_W^4}
          + \frac{3}{8} \frac{m_H^2 m_t^2}{m_W^4}
          + \frac{4}{9} \frac{m_Z^2 m_t^4}{m_W^4 m_H^2}
          + \frac{19}{12} \frac{m_Z^4 m_t^2}{m_W^4 m_H^2}
 \nonumber\\
&&{}
          - \frac{4}{9} \frac{m_t^4}{m_W^2 m_H^2}
          - \frac{20}{3} \frac{m_Z^2 m_t^2}{m_W^2 m_H^2}
          + \frac{35}{6} \frac{m_t^2}{m_H^2}
          \Bigg)
       + N_c n_G   \Bigg(
          - \frac{293}{1944}
          + \frac{143}{3888} \frac{m_Z^4}{m_W^4}
 \nonumber\\
&&{}
          + \frac{22}{27} \frac{m_Z^6}{m_W^4 m_H^2}
          - \frac{143}{1944} \frac{m_Z^2}{m_W^2}
          - \frac{44}{27} \frac{m_Z^4}{m_W^2 m_H^2}
          + \frac{40}{27} \frac{m_Z^2}{m_H^2}
          + \frac{4}{3} \frac{m_W^2}{m_H^2}
          \Bigg) \,,
\end{eqnarray}
for the top quark.
Here, all the masses and couplings are defined in the \MS scheme at
renormalization scale $\mu$, and
$n_G=3$ is the number of fermion generations.
Eqs.~(\ref{eq:z22}) and (\ref{eq:z21}) agree with
Ref.~\cite{Jegerlehner:2003py}.


\boldmath
\section{\MS mass of the bottom quark}
\label{app:mbMS}
\unboldmath

As anticipated at the end of Section~\ref{sec:four}, we present here a closed
expression for the ratio $m_b(\mu)/M_b$, in terms of on-shell
renormalized parameters.
At the two-loop level, we use the large-mass expansion with respect to the
top-quark mass.
Using the notation introduced in Section~\ref{sec:four}, we have
\begin{eqnarray}
\frac{m_b(\mu)}{M_b} &=& 1
       + \delta_{\rm QCD}(\mu)
      + \frac{\alpha}{4\pi} \Bigg\{
            N_c \frac{M_t^4}{M_W^2 M_H^2 S_w^2} \Big( 1 - \Lt \Big)
\nonumber\\
&&
          + \frac{M_t^2}{M_W^2 S_w^2} 
                \Big( -\frac{5}{16} + \frac{3}{8}\Lt \Big)
          + \frac{M_t^2 M_W^2}{(M_t^2-M_W^2)^2 S_w^2} 
                \Big( - \frac{3}{8} \Ltw \Big)
          + \frac{3}{8} \frac{M_W^2}{(M_t^2-M_W^2) S_w^2}
\nonumber\\
&&
          + a_v^2 \frac{M_Z^2}{M_H^2} \Big( - 4 + 12\Lz \Big)
          + \frac{M_H^2}{M_W^2 S_w^2} 
                \Big( -\frac{3}{8} - \frac{3}{8}\Lh \Big)
          + \frac{M_W^2}{M_H^2 S_w^2} 
                \Big( - \frac{1}{2} + \frac{3}{2} \Lw \Big)
\nonumber\\
&&
          + Q_b^2 \Big( - 4 + 3\Lb \Big)
          + v_b^2 \Big( - \frac{5}{2} + 3\Lz \Big)
          + a_v^2 \Big( - \frac{1}{2} - 3\Lz \Big)
\nonumber\\
&&
         + \frac{M_b^2}{M_W^2 S_w^2} \Bigg(
              \frac{11}{48}
            + \frac{1}{4} \Lb
            - \frac{1}{8} \Lt
            - \frac{1}{8} \Lz
            - \frac{3}{8} \Lh
\nonumber\\
&&
            + v_b^2 S_w^2 C_w^2 \Big( - \frac{8}{3} - 4 \Lb + 4 \Lz \Big)
            + O\left( \frac{M_b^4}{M_W^4} \right)
              \Bigg)
         \Bigg\}
\nonumber\\
&&
      + C_F \frac{\alpha_s(\mu)}{4\pi}\frac{\alpha}{4\pi} \Bigg\{
            N_c \frac{M_t^4}{M_W^2 M_H^2 S_w^2} 
               \Big( -2 + 16\Lt + 3\Lb - 3\Lb\Lt - 6\Lt^2 \Big)
\nonumber\\
&&
          + \frac{M_t^2}{M_W^2 S_w^2} 
               \Big( - \frac{13}{4} - \frac{3}{2}\Lt - \frac{15}{16}\Lb 
                     + \frac{9}{8}\Lb\Lt + \frac{9}{8}\Lt^2 \Big)
\nonumber\\
&&
          + \frac{1}{S_w^2} \frac{M_t^2 M_W^2}{(M_t^2-M_W^2)^2}
                   \Ltw \Big( \frac{33}{8} - \frac{9}{8}\Lb \Big)
          + \frac{1}{S_w^2} \frac{M_W^2}{M_t^2-M_W^2}
                   \Big( -\frac{51}{8} + \frac{9}{8}\Lb + \frac{3}{4}\Ltw \Big)
\nonumber\\
&&
          + \frac{1}{S_w^2} \Big( - \frac{63}{16} + \frac{3}{4}\Lt - 3\Lw \Big)
          + \frac{1}{S_w^2} \Bigg(
                   3\frac{M_t^2}{M_W^2}
                 + 3
                 + \frac{21}{4} \frac{M_W^2}{M_t^2-M_W^2}
\nonumber\\
&&
                 + \frac{9}{4} \frac{M_W^4}{(M_t^2-M_W^2)^2}
                  \Bigg) {\rm Li}_2\Big(1-\frac{M_W^2}{M_t^2}\Big)
          + \frac{M_H^2}{M_W^2 S_w^2}(1-\Lh)
               \Big( \frac{3}{2} - \frac{9}{8}\Lb \Big)
\nonumber\\
&&
          + a_b^2 \frac{M_Z}{M_H} (1-3\Lz)
                   \Big( 16 - 12\Lb \Big)
          + \frac{M_W^2}{M_H^2 S_w^2} (1-3\Lw)
                   \Big( 2 - \frac{3}{2}\Lb \Big)
\nonumber\\
&&
          + Q_b^2 \Big( \frac{7}{4} - 24\zeta_3 - 60\zeta_2 
                + 96 \zeta_2\log2 - 21\Lb + 9\Lb^2 \Big)
\nonumber\\
&&
          + v_b^2 \Big( \frac{23}{4} - \frac{15}{2}\Lb 
                 - 9\Lz + 9\Lb\Lz \Big)
          + a_b^2 \Big( \frac{55}{4} - \frac{3}{2}\Lb 
                 - 9\Lz - 9\Lb\Lz \Big)
         \Bigg\}       
\nonumber\\
&&
      + \left(\frac{\alpha}{4\pi}\right)^2 \Bigg\{   
          \left( N_c \frac{M_t^4}{M_W^2 M_H^2 S_w^2} \right)^2 
          \frac{(1-\Lt)(3+\Lt)}{2}
\nonumber\\
&&
      - \frac{N_c M_t^6}{16 M_W^4 M_H^2 S_w^4} (77 - 75 \Lt + 18 \Lt^2 + 8 \zeta_2)
      + \frac{N_c M_t^4}{M_H^4} \Bigg( 
          \frac{-7 + 5 \Lt + 3 \Lz + 3 \Lt \Lz}{4 C_w^4} 
\nonumber\\
&&
         + \frac{-7 + 5 \Lt + 3 \Lz + 3 \Lt \Lz}{2 C_w^2} 
         + \frac{3 (-7 + 5 \Lt + 2 \Lw + 2 \Lt \Lw + \Lz + \Lt \Lz)}{4 S_w^4} 
\nonumber\\
&&
         + \frac{-7 + 5 \Lt + 3 \Lz + 3 \Lt \Lz}{2 S_w^2} \Bigg)
\nonumber\\
&&
      + \frac{N_c M_t^4}{M_W^2 M_H^2} \Bigg( 
           \frac{67 - 67 \Lt + 32 \Lt^2 - 8 \Lz + 8 \Lt \Lz}{48 C_w^2} 
         + \frac{3 (-1 +\Lt)}{16 S_w^4} 
\nonumber\\
&&
         - \frac{-59 - 16 \Lb + 59 \Lt + 16 \Lb \Lt - 32 \Lt^2 + 24 \Lz - 24 \Lt \Lz}{48 S_w^2}
          \Bigg)
\nonumber\\
&&
      + \frac{N_c M_t^4}{M_W^4} 
          \frac{469 - 16 \Lh + 136 \Lt - 144 \Lh \Lt + 156 \Lt^2 + 8 \zeta_2}{128 M_W^4 S_w^4}
\nonumber\\
&&
      + \frac{M_t^4}{M_W^4 S_w^4} \Bigg( 
                -\frac{29}{512} - \frac{53}{128}\Lt + \frac{27}{128}\Lt^2 + \frac{1}{4} \zeta_2
              \Bigg) 
        + O(M_t^2)  \Bigg\} \,.
\label{runningmass}
\end{eqnarray}
Expanding the $O(\alpha\alpha_s)$ term of Eq.~(\ref{runningmass}) in powers of
$M_W^2/M_t^2$, we find agreement with Ref.~\cite{Kniehl:2004hfa}.

%
%

\end{document}